\documentclass[sigconf, nonacm]{acmart}

\newif\iflongversion
\longversiontrue

\usepackage[T1]{fontenc}
\usepackage[utf8]{inputenc}
\usepackage{booktabs}
\usepackage{multirow}
\usepackage{balance}

\usepackage[most]{tcolorbox}

\usepackage{xcolor}
\definecolor{shadecolor}{RGB}{192,192,192}

\usepackage{color}
\usepackage{verbatim}
\usepackage{graphicx}
\usepackage{balance}  
\usepackage{caption}
\usepackage{subcaption}
\usepackage{setspace}
\usepackage{float}
\usepackage{tikz}
\usepackage{amsmath}
\usepackage{breqn}
\usepackage{algorithm}
\usepackage[noend]{algpseudocode}
\usepackage{listings}

\usepackage[a-1b]{pdfx}
\usepackage{hyperref}

\definecolor{key-color}{rgb}{0.8, 0.47, 0.196}

\usepackage{enumitem}
\usepackage{hyphenat}

\usepackage{doi}
\usepackage{hyperref}
\expandafter\def\expandafter\UrlBreaks\expandafter{\UrlBreaks\do\/\do\-\do\.} 
\definecolor[named]{Purple}{cmyk}{0.55,1,0,0.15}
\definecolor[named]{DarkBlue}{cmyk}{1,0.58,0,0.21}
\hypersetup{bookmarksnumbered,unicode,naturalnames,colorlinks,breaklinks,
	linkcolor=Purple,
	citecolor=Purple,
	urlcolor=DarkBlue,
	filecolor=DarkBlue}

\graphicspath{ {images/} }

\newenvironment{squishedlist}
{
	\begin{list}{$\bullet$}
		{
			\setlength{\itemsep}{0pt}
			\setlength{\parsep}{1pt}
			\setlength{\topsep}{0.5pt}
			\setlength{\partopsep}{0pt}
			\setlength{\leftmargin}{01.0em}
			\setlength{\labelwidth}{1em}
			\setlength{\labelsep}{0.5em}
		}
	}
{
	\end{list}
}

\newenvironment{squishedenumerate}
{
	\begin{enumerate}[label=(\roman*),topsep=0.2em,itemsep=0.2em,leftmargin=1.6em] 
	}
	{
	\end{enumerate}
}

\newtheorem{example}{Example}

\newtheorem{desideratum}{Desideratum}

\newtheorem{guideline}{Guideline}

\newtheorem{ldbcQ}{}

\newtheorem{jobQ}{}

\tikzstyle{io} = [trapezium, trapezium left angle=70, trapezium right angle=110, minimum width=3cm, minimum height=0.7cm, inner sep=1pt, text badly centered, text width=1.2cm, draw=black, fill=blue!10]
\tikzstyle{decision} = [diamond, minimum width=1cm, minimum height=1cm, text width=1.5cm, text badly centered, inner sep=1pt, draw=black, fill=green!10]
\tikzstyle{process} = [rectangle, minimum width=3cm, minimum height=1cm, text width=2.5cm, text centered, draw=black, fill=red!10]
\tikzstyle{arrow}=[draw, -latex]

\graphicspath{ {./images/} }

\newcommand\vldbdoi{10.14778/3476249.3476297}
\newcommand\vldbpages{2491-2504}
\newcommand\vldbvolume{14}
\newcommand\vldbissue{11}
\newcommand\vldbyear{2021}
\newcommand\vldbauthors{\authors}
\newcommand\vldbtitle{\shorttitle} 
\newcommand\vldbavailabilityurl{https://github.com/graphflow/graphflow-columnar-techniques}
\newcommand\vldbpagestyle{empty}

\begin{document}
\title{Columnar Storage and List-based Processing for Graph Database Management Systems}

\author{Pranjal Gupta, Amine Mhedhbi, Semih Salihoglu}
\affiliation{%
	\institution{University of Waterloo}
}
\email{{pranjal.gupta, amine.mhedhbi, semih.salihoglu}@uwaterloo.ca}

\begin{abstract}
We revisit column-oriented storage and query processing techniques in the context of contemporary graph database management  systems (GDBMSs). 
Similar to column-orien\-ted RDBMSs, 
GDBMSs support read-heavy analytical workloads that however have 
fundamentally different data 
access patterns than traditional analytical workloads. 
We first derive a set of desiderata 
for optimizing storage and query processors of GDBMS based on their 
access patterns. We then present the design of 
columnar storage, compression, and query processing techniques based on these desiderata.
In addition to showing direct integration of existing techniques from
columnar RDBMSs, we also propose novel ones that are optimized for GDBMSs. 
These include a novel list-based query processor, 
which avoids expensive data copies of traditional block-based processors under many-to-many joins,
a new data structure we call single-indexed edge property pages and 
an accompanying edge ID scheme, 
and a new application of Jacobson's bit vector index for compressing NULL values and empty lists. 
We integrated our techniques into the GraphflowDB in-memory GDBMS. 
Through extensive experiments, we demonstrate the scalability 
and query performance benefits of our techniques. 
\end{abstract}

\maketitle

\pagestyle{\vldbpagestyle}

\section{Introduction}
\label{sec:introduction}

Contemporary GDBMSs are data management software such as Neo4j \cite{neo4j}, Neptune~\cite{neptune}, TigerGraph~\cite{tigergraph}, and GraphflowDB~\cite{kankanamge:graphflow, mhedhbi:sqs} that adopt the property graph data model~\cite{neo4j-property-graph-model}. In this model, application data is represented as a set of vertices and edges, which represent the entities and their 
relationships, and key-value properties on the vertices and edges. GDBMSs support a wide range of analytical applications, such as fraud detection and recommendations in financial, e-commerce, or social networks~\cite{sahu:survey} that search for patterns in a graph-structured database, which 
require reading large amounts of data. In the context of RDBMSs, column-oriented systems~\cite{oracle-col, monet-2decades, c-store, boncz-vectorwise} employ a set of read-optimized storage, indexing, and query processing techniques to support traditional analytical applications, such as business intelligence and reporting, that also process large amounts of data. As such, these techniques are relevant for improving the performance and scalability of GDBMSs. 

In this paper, we revisit columnar storage and query processing techniques in the context of GDBMSs. 
Specifically, we focus on an in-memory GDBMS setting and discuss the applicability of columnar storage and compression techniques for storing different components of graphs~\cite{abadi-sparse-col, abadi-col-comp, c-store, boncz-comp}, 
and block-based query processing~\cite{col-vs-row, boncz-monet-vectorized}. 
Despite their similarities, workloads in GDBMSs and columnar RDBMSs also have fundamentally 
different access patterns.
For example,
workloads in GDBMSs contain large many-to-many joins, which are not frequent in column-oriented RDBMSs.
This calls for redesigning columnar techniques in the context of GDBMSs. The contributions of this paper are as follows. 

\vspace{2pt}
\noindent {\em \textbf{Guidelines and Desiderata:}}  We begin in Section~\ref{sec:guidelines} by analyzing the properties of data access patterns in GDBMSs. For example, we observe that different components of data stored in GDBMSs can have some structure and the order in which operators access vertex and edge properties often follow the order of edges in adjacency lists. This analysis instructs a set of guidelines and desiderata for designing the physical data layout and query processor of a GDBMS. 

\vspace{2pt}
\noindent  {\em \textbf{Columnar Storage:}} Section~\ref{sec:columnar-storage} explores the application of columnar data 
structures for storing different data components in GDBMSs. While existing columnar 
structures can directly be used for storing vertex properties and many-to-many 
(n-n) edges, we observe that using straightforward edge columns, 
to store properties of n-n edges does not guarantee sequential access 
when reading edge properties in either forward or backward directions.
An alternative, which we call {\em double-indexed property CSRs}, can achieve sequential access 
in both directions but requires duplicating edge properties.   
We then describe an alternative design point, 
{\em single-directional property pages}, that avoids duplication and 
achiev\-es good locality when reading properties of edges in one direction and still
guarantees random access in the other. This requires using a new edge ID scheme
that is conducive to extensive compression when storing them 
in adjacency lists without any decompression overheads. 
Lastly, as a new application of vertex columns, we show that single cardinality edges and edge properties, i.e. those with one-to-one (1-1), one-to-many (1-n) or many-to-one (n-1) cardinalities, are stored more efficiently with vertex  columns instead of the structures we describe for n-n edges.

\vspace{2pt}
\noindent  {\em \textbf{Columnar Compression:}} In Section~\ref{sec:columnar-compression}, we review existing columnar 
compression techniques, such as dictionary encoding, that satisfy our desiderata and can be directly applied to 
GDBMSs. We next show that existing techniques for compressing
NULL values in columns from references~\cite{abadi-sparse-col, abadi-col-comp} by Abadi et al. 
lead to very slow accesses to arbitrary non-NULL values.
We then review Jacobson's bit vector index~\cite{jacobson:bitvector, jacobson:thesis} to support
constant time rank queries, which has found several prior applications
e.g., in a range filter structure in databases~\cite{zhang:surf}, in information retrieval~\cite{gonnet:ir, navarro:indexes} and computational geometry~\cite{bose:geometry, navarro:geometry}. 
We show how to enhance one of Abadi's schemes with an adaptation of Jacobson's index 
to  provide constant-time access to arbitrary non-NULL values,
with a small increase in storage overhead compared to prior techniques.
	
\vspace{2pt}
\noindent  {\em \textbf{List-based Processing:}} 
In Section~\ref{sec:list-based-processing}, 
we observe that traditional 
block-based processors or columnar RDBMSs~\cite{col-vs-row, boncz-vectorwise1} 
process fixed-length blocks of data in
tight loops, which achieves good CPU and cache utility but results in
expensive data copies under n-n joins.
To address this, we propose 
a new block-based processor we call {\em list-based processor (LBP)}, which modifies
traditional block-based processors in two ways to tailor them for GDBMSs: (i) Instead of representing 
the intermediate tuples processed by operators as a single group of equal-sized 
blocks, we represent them as multiple factorized groups of blocks. We call these {\em list groups}. 
LBP avoids expensive data copies 
by flattening blocks of some groups into single values when performing
n-n joins.  
(ii) Instead of fixed-length blocks, LBP uses variable length blocks that take the 
lengths of adjacency lists that are represented in the intermediate tuples. 
Because adjacency lists are already stored in memory consecutively, 
this allows us to avoid materializing adjacency lists during join processing, improving query performance. 

We integrated our techniques into GraphflowDB~\cite{kankanamge:graphflow}. 
We present extensive experiments that demonstrate the scalability and performance 
benefits (and tradeoffs) of our techniques
both on microbenchmarks and on the LDBC and JOB benchmarks against 
a row-based Volcano-style implementation of the system, an open-source version of a 
commercial GDBMSs, and two column-oriented RDBMSs.
Our code, queries, and data are available here~\cite{graphflowdb-github}.

\section{Background} 
\label{sec:background}

\begin{figure}
	\centering
	\includegraphics[width=1\columnwidth]{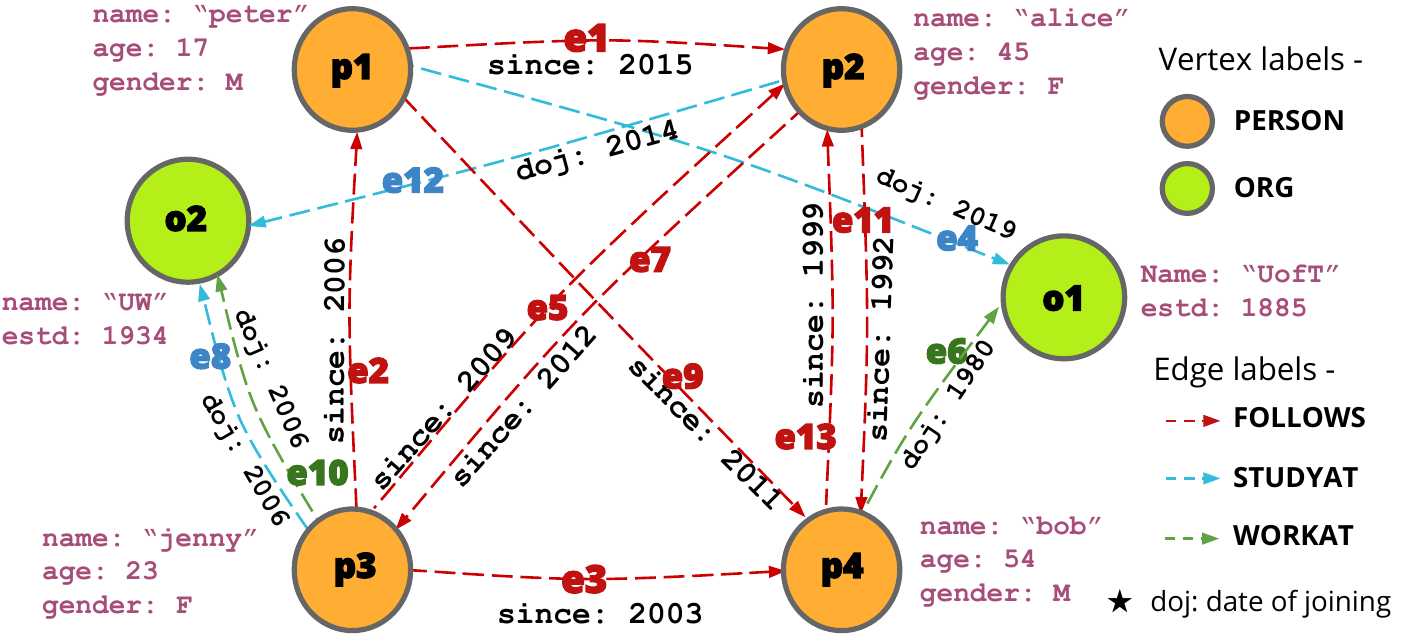}
	\vspace{-17pt}
	\caption{Running example graph.}
	\label{fig:runn}
	\vspace{-5pt}
\end{figure}

In the property graph model, vertices and edges have labels and arbitrary key value properties.
Figure~\ref{fig:runn} shows a property graph that will serve as our running example, which 
contains vertices with \texttt{PERSON} and \texttt{ORGANIZATION (ORG)} labels, and edges with \texttt{FOLLOWS}, \texttt{STUDYAT} and \texttt{WORKAT} labels.

There are three storage components of GDBMSs: (i) topology, i.e., adjacencies of vertices; (ii) vertex properties;
and (iii) edge properties. In every native GDBMS we are aware of, the topology is stored in
data structures that organize data in \emph{adjacency lists}~\cite{bonifati-adj-lists}, such as in
compressed sparse row (CSR) format. 
Typically, given the ID of a vertex $v$, the system can in constant-time access $v$'s adjacency list,
 which contains a list of (edge ID, neighbour ID) pairs. Typically, the adjacency 
list of $v$ is further clustered by edge label which enables efficient traversal of the neighbourhood of $v$, given a 
particular label. Vertex and edge properties can be stored in a number of ways. For example, some systems use a key-value store, such as DGraph~\cite{dgraph} and JanusGraph~\cite{janusgraph}, and some use a variant of {\em interpreted attribute layout}~\cite{beckmann:sparse}, where records consist of
variable-sized key-value properties. 
Records can be located consecutively in disk or memory or have pointers to each other, as in Neo4j. 

Queries in GDBMSs consist of a subgraph pattern $Q$ that describes the joins 
in the query (similar to SQL's FROM) and optionally predicates on these patterns with final 
group-by-and-aggregation operations. We assume a GDBMS with a query processor that
uses variants of the following relational operators, which is the case in many GDBMSs, e.g., Neo4j~\cite{neo4j}, Memgraph~\cite{memgraph}, or GraphflowDB:

\noindent \texttt{Scan}: Scans a set of vertices from the graph.
	
\noindent \texttt{Join} (e.g. \texttt{Expand} in Neo4j and Memgraph, \texttt{Extend} in GraphflowDB): 
Performs an index nested loop join using the adjacency list index to match an edge of $Q$.   
Takes as input a partial match $t$ that has matched $k$ of the query edges in $Q$. For each $t$, 
\texttt{Join} extends $t$ by matching an unmatched query edge $qv_s$$\rightarrow$$qv_d$, where $qv_s$ or 
$qv_d$ has already been matched. For example if $qv_s$ has already been matched to data vertex $v_i$, 
then the operator produces one $(k+1)$-match for each edge-neighbour pair in $v_i$'s forward
adjacency list\footnote{
GraphflowDB can perform an intersection of 
multiple adjacency lists if the pattern is cyclic (see reference~\cite{mhedhbi:sqs}).}.

\noindent \texttt{Filter}: Applies a predicate $\rho$ to a partial match $t$,
reading any necessary vertex and edge properties from storage.

\noindent \texttt{Group By And Aggregate}: Performs a standard group by and aggregation computation
on a partial match $t$.	

\begin{figure}
	\centering
	\includegraphics[scale=0.70]{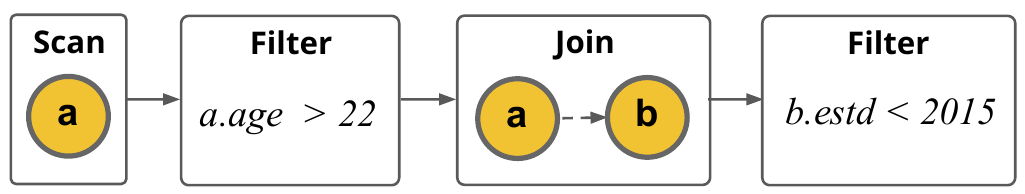}
	\vspace{-5pt}
	\caption{Query plan for the query in Example~\ref{ex:cypher-example}.}
	\label{fig:ex-qp}
	\vspace{-5pt}
\end{figure}

\vspace{-4pt}
\begin{example}
	\label{ex:cypher-example}
	Below is an example query written in the Cypher language~\cite{cypher}:
	{\em 
		\begin{lstlisting}[numbers=none,  showstringspaces=false,belowskip=0pt ]
MATCH (a:PERSON)$-$[e:WORKAT]$\rightarrow$(b:ORG)
WHERE a.age $>$ 22 AND b.estd < 2015 RETURN *\end{lstlisting}
}
\vspace{2pt}
	\noindent The query returns all persons $a$ and their workplaces $b$, where $a$ is older than
	22 and $b$ was established before 2015. Figure~\ref{fig:ex-qp} shows a typical plan for this query.
\end{example}

\section{Guidelines and Desiderata} 
\label{sec:guidelines}

We next outline a set of guidelines and desiderata for organizing the physical data layout and query processor
of GDBMSs.
We assume edges are doubly-indexed in 
forward and backward adjacency lists, as in every GDBMS we are aware of.
We will not optimize this duplication as this is needed
for fast joins from both ends of edges.

\begin{guideline}
\label{gdl:edge-vertex-order}
Edge and vertex properties are read in the same order as the edges appear in adjacency lists after joins.
\end{guideline}
\vspace{-2pt}
\noindent Observe that \texttt{JOIN} accesses the edges and neighbours of a vertex $v_i$ in the order 
these edges appear in $v_i$'s adjacency list $L_{v_i}=\{(e_{i1}, v_{i1}) ..., (e_{i\ell}, v_{i\ell})\}$.
 If the next operator  also needs to access the 
properties of these edges or vertices, e.g., \texttt{Filter} in Figure~\ref{fig:ex-qp}, these accesses
will be in the same order. 
Our first desiradata is to store the properties of $e_{i1}$ to $e_{i\ell}$ sequentially in the same order.
Ideally, a system should also store the properties $v_{ij}$ sequentially in the same order
but in general this would require
prohibitive data replication because while each $e_{ij}$ appears in two adjacency lists,
each $v_{ij}$ appears in as many lists as the degree of $v_{ij}$.

\vspace{-4pt}
\begin{desideratum}
\label{des:edge-ordered-reads}
Store and access the properties of edges sequentially in the order edges appear in adjacency lists.
\end{desideratum}

\begin{guideline}
\label{gdln:fast-decompress}
Access to vertex properties will not be to sequential locations and many adjacency lists are very small.
\end{guideline}
\vspace{-4pt}
\noindent Guideline~\ref{gdl:edge-vertex-order} implies that we should expect random accesses in memory when an
operators access vertex properties.
In addition, real-world graph data with n-n relationships have
power-law degree distributions~\cite{leskovec:laws}. So, there are often many short adjacency lists in the dataset.
For example, the \texttt{FLICKR}, \texttt{WIKI} graphs that we use, have single edge labels with average degrees
of only 14 and 41, and the Twitter dataset used in many prior work on GDBMSs~\cite{konect} has
a degree of 35.
Therefore when processing queries with two or more joins, 
reading different adjacency lists will require iteratively reading
a short list followed by a random access.
This implies that techniques that require decompressing blocks of data, say a few KBs, 
to only read a single vertex property or a single short adjacency list can be prohibitively expensive.

\vspace{-4pt}
\begin{desideratum}
\label{des:compression}
If compression is used, decompressing arbitrary data elements in a compressed block should happen in constant time.
\end{desideratum}  

\begin{guideline}
\label{gdln:graph-schema}
Graph data often has partial structure.
\end{guideline}
\vspace{-4pt}
\noindent Although the property graph model is semi-structured, 
data in GDBMSs often have some structure. 
One reason for this is because the data in GDBMSs 
sometimes comes from structured data from RDBMSs as observed in a recent user survey \cite{sahu:survey}.
In fact,  several vendors and academics are actively working on defining a schema language for 
property graphs \cite{schema-validation-bonifati, defining-schema-hartig}.
Common structure are: 

\begin{squishedenumerate}
		
\item {\em Edge label determines source and destination vertex labels.} 
For example, in the popular LDBC social network benchmark (SNB), 
\texttt{KNOWS} edges exist only between vertices of label \texttt{PERSON}.
	
\item {\em Label determines vertex and edge properties.} Similar to attributes of a relational table, properties on 
an edge or vertex and their datatypes can often be determined by the label. 
For example, this is the case for every vertex and edge label in LDBC.

\item[(iii)] {\em Edges with single cardinality.} Edges might have cardinality constraints:
1-n (single cardinality in the backward edges), n-1 (single cardinality in the forward edges), 1-1, and n-n. 
An example of 1-n cardinality from LDBC SNB is that each \texttt{organization} has one \texttt{isLocatedIn} edge.

\end{squishedenumerate}
We refer to edges that satisfy properties (i) and (ii) as {\em structured edges}
and properties that satisfy property (ii) as {\em structured vertex/edge property}.
Other edges and properties will be called {\em unstructured}. 
The existence of such structure in some graph data motivates 
our third desideratum:
	
\vspace{-4pt}
\begin{desideratum}
\label{des:structure}
Exploit structure in the data for space-efficient storage and faster access to data.
\end{desideratum}

\section{Columnar Storage}
\label{sec:columnar-storage}

\begin{table}[t!]
	\centering
	\bgroup
	\setlength{\tabcolsep}{5pt}
	\def\arraystretch{1.3}
	\captionsetup{justification=centering}
	\caption{Columnar data structures and data components they are used for. V-Column stands for vertex column.}
		\label{tbl:summ}
	\vspace{-5pt}
	\begin{tabular}{ |c|p{5.4cm}| } 
		\hline
		\textbf{Data} & \textbf{Columnar data structure} \\
		\hline
		\multirow{1}{*}{Vertex Properties} & V-Column \\
		\hline 
		\multirow{3}{*}{Edge Properties} & V-Column: of src when n-1, of dst when 1-n, of either src or dst when 1-1\\
		\cline{2-2}
		& Single-indexed prop. pages when n-n\\
		\hline 
		\multirow{1}{*}{Fwd Adj. lists} & V-Column when 1-1 and n-1, CSR o.w. \\
		\hline 
		Bwd Adj. lists & V-Column when 1-1 and 1-n, CSR o.w. \\
		\hline 
	\end{tabular}
	\egroup
\end{table}

We next explore using columnar structures for storing data in GDBMSs to meet the desiderata from Section~\ref{sec:guidelines}. 
For reference, Table~\ref{tbl:summ} presents the summary of the columnar structures we use and the data they store. 
We start with directly applicable structures and then describe our new single-indexed property pages 
structure and its accompanying edge ID scheme to store edge properties.

\vspace{-4pt}
\subsection{Directly Applicable Structures}
\label{subsec:storage-directly-applicable}

\subsubsection{CSR for n-n Edges}
\label{subsubsec:csr-storage}

\begin{figure}
	\hfill\includegraphics[scale=0.64]{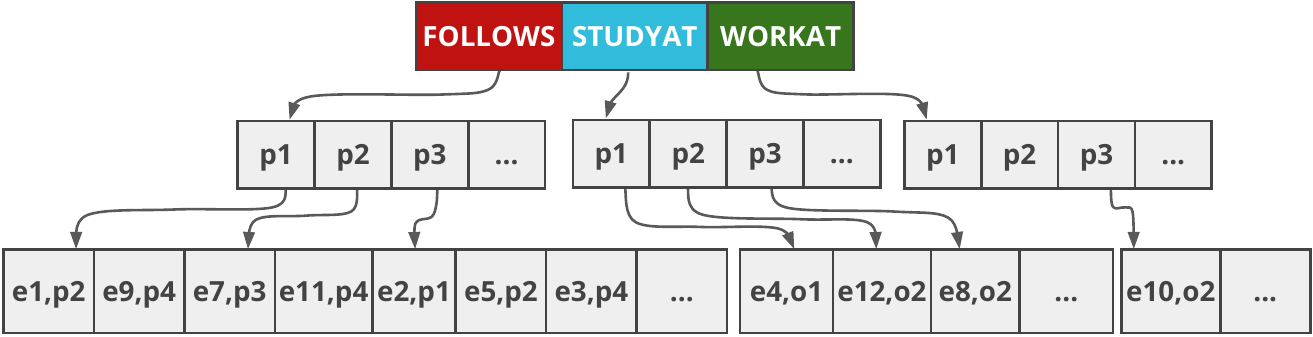}\hspace*{\fill}
	\vspace{-4pt}
	\captionsetup{justification=centering}
	\caption{Example forward adjacency lists implemented as a 2-level CSR structure for the example graph.}
	\vspace{-3pt}
	\label{fig:adjlists}
\end{figure}

CSR is an existing columnar structure that is
widely used by existing GDBMSs to store edges. A CSR, shown in Figure~\ref{fig:adjlists}, effectively stores 
a set of (vertex ID, edge ID, neighbour ID) triples sorted by vertex ID, where the 
vertex IDs are compressed similar to run-length encoding.
In this work, we store the edges of each edge label with n-n cardinality in a separate CSR.
As we discuss next, we can store the edges with other cardinalities more efficiently than a CSR by using vertex columns.

\subsubsection{Vertex Columns for Vertex Properties, Single Cardinality Edges and Edge Properties}
\label{subsub:vertex-columns}
 With an appropriate vertex ID scheme, columns can be directly used for storing 
structured vertex properties in a compact manner.  
Let $p_{i,1},  p_{i,2}, ... p_{i, n}$ be the structured vertex properties of
vertices with label $lv_i$. We have a \emph{vertex column} for each $p_{i,j}$, that stores $p_{i,j}$ properties of vertices in consecutive
locations. 
Then we can adopt a (vertex label, label-level positional offset) ID scheme 
and ensure that offsets with the same label are consecutive.
As we discuss in Section~\ref{sec:storage-optimizations}, this ID scheme also 
can be compressed by factoring out vertex labels.

\begin{figure}
	\hfill\includegraphics[scale=0.58]{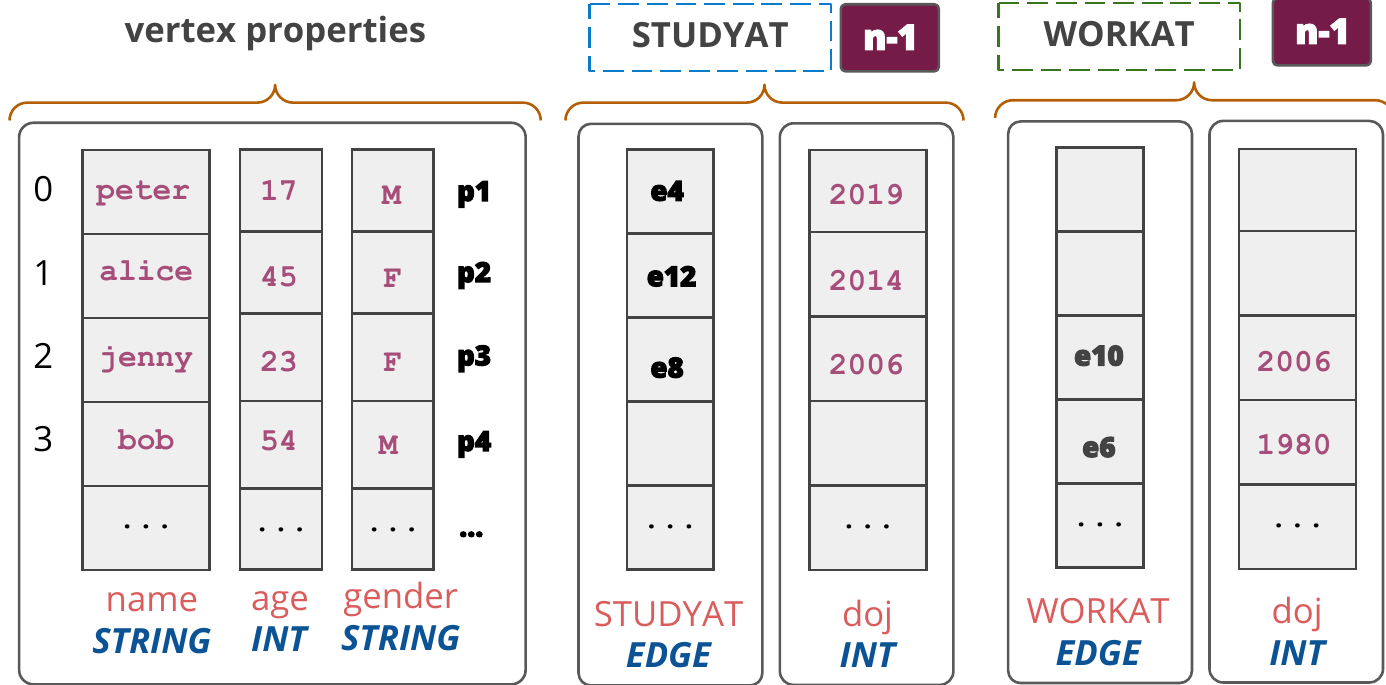}\hspace*{\fill}
	\captionsetup{justification=centering}
	\caption{Example vertex columns storing vertex properties and single-cardinality edges and their properties.}
    \vspace{-10pt}
	\label{fig:single-cardinality-cols}
\end{figure}

Similarly, we can store single cardinality edges, i.e., those with 1-1, 1-n, or n-1 constraints, 
and their properties 
directly
as a \emph{property} of source or destination vertex of the edges in a vertex column
and directly access them using a vertex positional offset.
As we momentarily discuss, 
this is more efficient both in terms of storage and access time 
than the structures we cover for storing properties of n-n edges (Desideratum~\ref{des:structure}).
Figure~\ref{fig:single-cardinality-cols} shows single cardinality STUDYAT and WORKAT edges from our example and their properties stored as
vertex column of \texttt{PERSON} vertices.

\subsection{Single-indexed Edge Property Pages for Properties of n-n Edges}
\label{sec:edge-property-columns}

Recall Desideratum~\ref{des:edge-ordered-reads} that 
access to edge properties should be in the same order of the edges in adjacency lists.
We first review two columnar structures,
{\em edge columns} and  {\em double-indexed property CSRs}, the former of which 
has low storage cost but does not satisfy Desideratum~\ref{des:edge-ordered-reads} and the latter
has high storage cost but satisfies Desideratum~\ref{des:edge-ordered-reads}. 
We then describe a new design, which we call {\em single-indexed property pages}, 
which has low storage cost as edge columns and with a new edge ID scheme can partially 
satisfy Desideratum~\ref{des:edge-ordered-reads}, so dominates 
edge columns in this design space.

\noindent {\bf Edge Columns:}
We can use a separate {\em edge column} for each 
property $q_{i, j}$ of edge label $le_i$. Then with an appropriate
edge ID scheme, 
such as (edge label, label-level positional offset), 
one can perform a random access to read the $q_{i, j}$ property of an edge $e$. 
This design has low storage cost and stores each property once but does not
store the properties according to any order.
In practice, the order would be determined by the sequence of edge insertions and deletions.

\noindent {\bf Double-Indexed Property CSRs.}
An alternative is to mimic the storage of adjacency lists in the CSRs in 
separate CSRs that store edge properties. 
For each vertex $v$
we can store $q_{i,j}$ twice in \emph{forward} and {\em backward property lists}. 
This design provides sequential read of properties in both directions, thereby satisfying 
Desideratum~\ref{des:edge-ordered-reads}, but also requires double the storage of edge columns. 
This can often be prohibitive especially for in-memory systems, as many graphs 
have orders of magnitude more edges than vertices.

A natural question is: {\em Can we avoid duplicate
storage of double-indexed property CSRs but still achieve sequential reads?} We
next show a structure that with an appropriate edge ID scheme obtains sequential reads in one direction,
so partially satisfying Desideratum~\ref{des:edge-ordered-reads}. This structure therefore dominates 
edge columns in design. 

\noindent {\bf Single-indexed property pages:}
A first natural design 
uses only one property CSR, say forward. We call this structure
{\em single-indexed property CSR}.
Then, properties 
can be read sequentially in the forward direction.
However, reading a property in the other direction quickly, specifically with 
constant time access, requires a new edge ID scheme. 
To see this 
suppose a system has read the backward adjacency lists of a vertex $v$ with label $le_i$, 
$\{(e_1, nbr_1), ...,$ $(e_k, nbr_k)\}$, and needs to 
read the $q_{i, j}$ property of these edges. 
Then given say $e_1$, we need to be able to
read $e_1$'s $q_{i, j}$ property from the forward property list $P_{nbr_1}$ of $nbr_1$. 
With a standard
edge ID scheme, for example one that assigns consecutive IDs to all edges with label $le_i$,
the system would need to first find the offset $o$ of $e_1$ in forward adjacency list of $nbr_1$, $L_{nbr_1}$,
which may require scanning the entire $L_{nbr_1}$, which is not constant time. 

Instead, we can adopt a 
new edge ID scheme that stores the following: 
(edge label, source vertex ID, list-level positional offset)\footnote{If we use
the backward property CSR, the second component would instead be the destination vertex ID.}. 
With this scheme a system can: (i) identify each edge, e.g., perform equality checks
between two edges; and (ii) read the offset $o$ directly from edge IDs, so
reading edge properties in the opposite direction (backward in our example)
can now be constant time.  In addition, this scheme
can be more space-efficient than schemes that assign 
consecutive IDs to all edges as 
its first two components can often be compressed (see Section~\ref{sec:storage-optimizations}).
However, single-indexed property CSR and this edge ID scheme 
has two limitations.
First access to properties in the `opposite direction' requires two random accesses, 
e.g., first access obtains the $P_{nbr_1}$ list using $nbr_1$'s ID and the second access 
reads a $q_{i, j}$ property from $P_{nbr_1}$. Second, although we do not focus on updates in this paper, 
using edge IDs that contain positional offsets 
has an important consequence for GDBMSs. 
Observe that positional offsets that are used by GDBMSs are explicitly stored in data structures.
\iflongversion
For example, in every native GDBMS we are aware of vertex IDs are used as positional
offsets to read vertex properties and they are also explicitly stored in adjacency lists.
This is unlike traditional columnar RDBMSs, where positional offsets,
specifically row IDs of tuples, are implicit and not explicitly stored. Therefore,
when deletions happen, GDBMSs initially leave gaps in their data structures,
and recycle deleted IDs when insertions happen. For example,
Neo4j's nodestore.db.id file keeps track of deleted IDs for later recycling~\cite{neo4j-deletes}.
Similarly, the list-level positional offsets 
need to be recycled. 
\fi
This may leave many gaps in adjacency lists
because to recycle a list-level offset, the system needs to wait for another insertion into the 
same adjacency list, which may be infrequent.

\begin{figure}
	\hfill\includegraphics[scale=0.75]{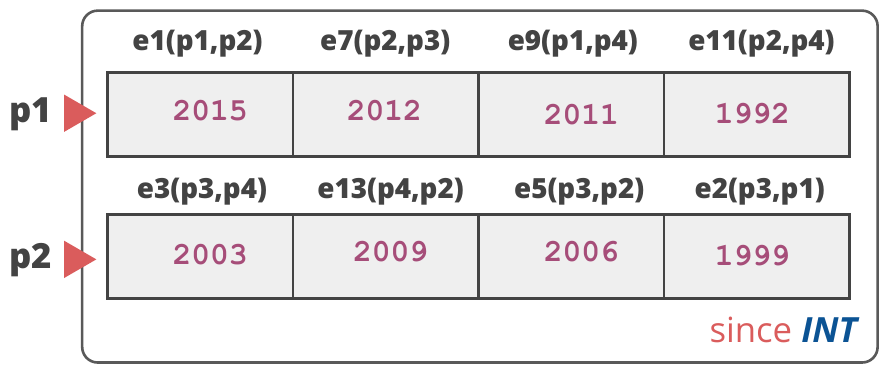}\hspace*{\fill}
	\captionsetup{justification=centering}
	\vspace{-5pt}
	\caption{Single-indexed property pages for \texttt{since} property of \texttt{FOLLOWS} edges in the example graph. $k=2$.}
	\vspace{-10pt}
	\label{fig:paged}
\end{figure}

Our {\em single-indexed property pages} 
addresses these two issues (Figure~\ref{fig:paged}). 
We store $k$ property lists (by default 128) in a property page. 
In a property page, properties of the same list does not have to be consecutively. However, because we use a small value of $k$,
these properties are stored in close-by memory locations. 
We modify the edge ID scheme above to use page-level positional offsets. This 
has two advantages. First, given a positional offset, the system can directly read 
an edge property (so we avoid the access to read $P_{nbr_1}$). 
Second, the system can recycle a page-level offset whenever any one of the k
lists get a new insertion. For reference, Figure~\ref{fig:paged} shows the single-indexed 
property pages in the forward direction for \texttt{since} property of edges with label \texttt{FOLLOWS} when $k$=2.

\section{Columnar Compression}
\label{sec:columnar-compression}

Compression and query processing on compressed data are widely 
used in columnar RDBMSs. 
We start by reviewing techniques that apply directly to GDBMSs and are not novel. 
We then discuss the cases when we can compress the new vertex and edge ID schemes from
Section~\ref{sec:columnar-storage}. Finally, we review existing NULL compression
schemes from columnar RDBMSs ~\cite{abadi-sparse-col, abadi-col-comp}
and enhance one of them with Jacobson's bit vector index
to make it suitable for GDBMSs.

\vspace{-5pt}
\subsection{Directly Applicable Techniques}
\label{sec:col-existing}

Recall our Desideratum~\ref{des:compression} that because access to vertex properties cannot be localized
and because many adjacency lists are very short, 
the compression schemes that are suitable for in-memory GDBMSs need to either avoid decompression
completely or support decompressing arbitrary elements in a block in constant time.
This is only possible if the elements are encoded in \emph{fixed-length codes} instead of variable-length codes.  We 
review dictionary encoding and leading 0 suppression, which we integrated in our 
implementation and refer readers to references~\cite{abadi-col-comp, goldstein:for, lemire:integer} 
for details of other fixed-length schemes, such as frame of reference.

\noindent \textbf{Dictionary encoding:} 
This is perhaps the most common encoding scheme to be used 
in RDBMSs~\cite{abadi-col-comp, boncz-comp, westmann-comp}.
This scheme maps a domain of values into compact 
 codes using a variety of schemes \cite{abadi-col-comp, dat-comp, boncz-comp}, some 
producing variable-length codes, such as Huffmann encoding, and others fixed-length codes~\cite{abadi-col-comp}. 
We use dictionary encoding to map a categorical edge or vertex property $p$, e.g., gender property 
of \texttt{PERSON} vertices in LDBC SNB dataset, that takes on $z$ different values to $\lceil log_2(z)/8 \rceil$ bytes 
(we pad $log_2(z)$ bits with 0s to have a fixed number of bytes).

\noindent \textbf{Leading 0 Suppression:} This scheme omits storing leading zero bits in each value of a given block of data \cite{beckmann:sparse}. 
We adopt a fixed-length variant of this for storing components of edge and vertex IDs, e.g., if the maximum size of a property page is $t$, we use $\lceil log_2(t)/8\rceil$ many bytes for the page-level positional offset of edge IDs. 

\vspace{-5pt}
\subsection{Factoring Out Edge/Vertex ID Components}
\label{sec:storage-optimizations}

Our vertex and edge ID schemes from Sections~\ref{sec:columnar-storage} decompose 
the IDs into many small components, which can be factored out when the data depicts some structure (Desideratum~\ref{des:structure}). This allows compression without the need to decompress while scanning.
Recall that the ID of an edge $e$ is a triple (edge label, source/destination vertex ID, page-level positional offset) and 
the ID of a vertex $v$ is a pair (vertex label, label-level positional offset). 
Recall also that GDBMSs store (edge ID, neighbour ID) pairs in adjacency lists. 
First, the vertex IDs inside the edge ID can be omitted because 
this is the neighbour vertex ID, which is already stored in the pairs. 
Second
edge labels can be omitted because we cluster our adjacency lists by edge label. The only 
components that need to be stored are: (i) positional offset of the edge ID; 
and (ii) vertex label and positional offset of neighbour vertex ID. 
When the data depicts some structure, we can further factor out some of these components as follows: 

\begin{squishedlist}
\item {\em Edges do not have properties:} Often, edges of a particular label do not have any 
properties and only represent the relationships
between vertices. For example, 10 out of 15 edge labels in LDBC SNB
do not have any properties. In this case, 
edges need not be identifiable, as the 
system will not access their properties. We can therefore distinguish two edges by their 
neighbour ID and edges with the same IDs are simply replicas of each other. Hence, 
we can completely omit storing the positional offsets of edge IDs.

\item {\em Edge label determines neighbour vertex label.} 
Often, edge labels in the graph are between a single source and destination vertex label,
e.g., \texttt{Knows} edges in social networks are between \texttt{Person} nodes.
In this case, we can omit storing the vertex label of the neighbour ID. 

\item {\em Single cardinality edges:} Recall from Section~\ref{subsub:vertex-columns} that the properties for single cardinality edges can be stored in vertex columns. So we can directly read these properties by using the source or destination vertex ID. So, the page-level positional offsets of these edges can be omitted. 
	
\end{squishedlist}

\begin{figure}
	\hfill\includegraphics[scale=0.60]{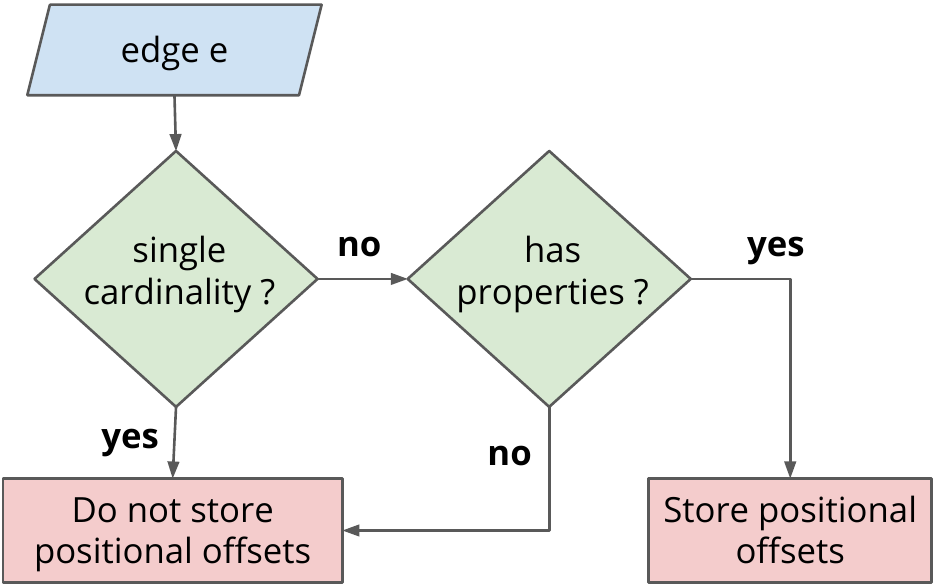}\hspace*{\fill}
	\captionsetup{justification=centering}
	\caption{Decision tree for storing page-level positional offsets of edges in adjacency lists.}
	\label{fig:dec2}
	\vspace{-10pt}
\end{figure}

\noindent Figures~\ref{fig:dec2} shows our decision tree to decide when to omit storing the page-level positional offsets in edge IDs.

\subsection{NULL and Empty List Compression}
\label{sec:null}

Edge and vertex properties can often be quite sparse in real-world graph data. 
Similarly, many vertices can have empty adjacency lists in CSRs. Both can be seen as different columnar structures containing NULL values. 
Abadi in reference~\cite{abadi-sparse-col} describes a design space
of optimized techniques for compressing  NULLs in columns.
All of these techniques list non-NULL elements consecutively in a `non-NULL values column' and use a secondary structure  to indicate the positions of these non-NULL values. 
The first technique in Abadi's paper, lists positions of each non-NULL value consecutively, which is suitable for very sparse columns, e.g., with $>90\%$ NULLs.
Second, for dense columns, lists non-NULL values as a sequence of pairs, 
each indicating a range of positions with non-NULL values. Third, for columns with intermediate sparsity, 
uses a bit vector to indicate if each location is NULL or not. The last technique is quite compact 
and requires only 1 extra bit  per each element in a column.

However, none of these techniques are directly applicable to GDBMSs as they do not allow constant-time access to non-NULL values (Desideratum~\ref{des:compression}). 
To support 
constant-time access to a non-NULL value at position $p$, the secondary
structure needs to support two operations in constant time: (i) check if $p$ is NULL or not; and (ii) if it is non-NULL, compute the {\em rank} of $p$, i.e., the number of non-NULL values before $p$.

\begin{figure}
	\hfill\includegraphics[scale=0.58]{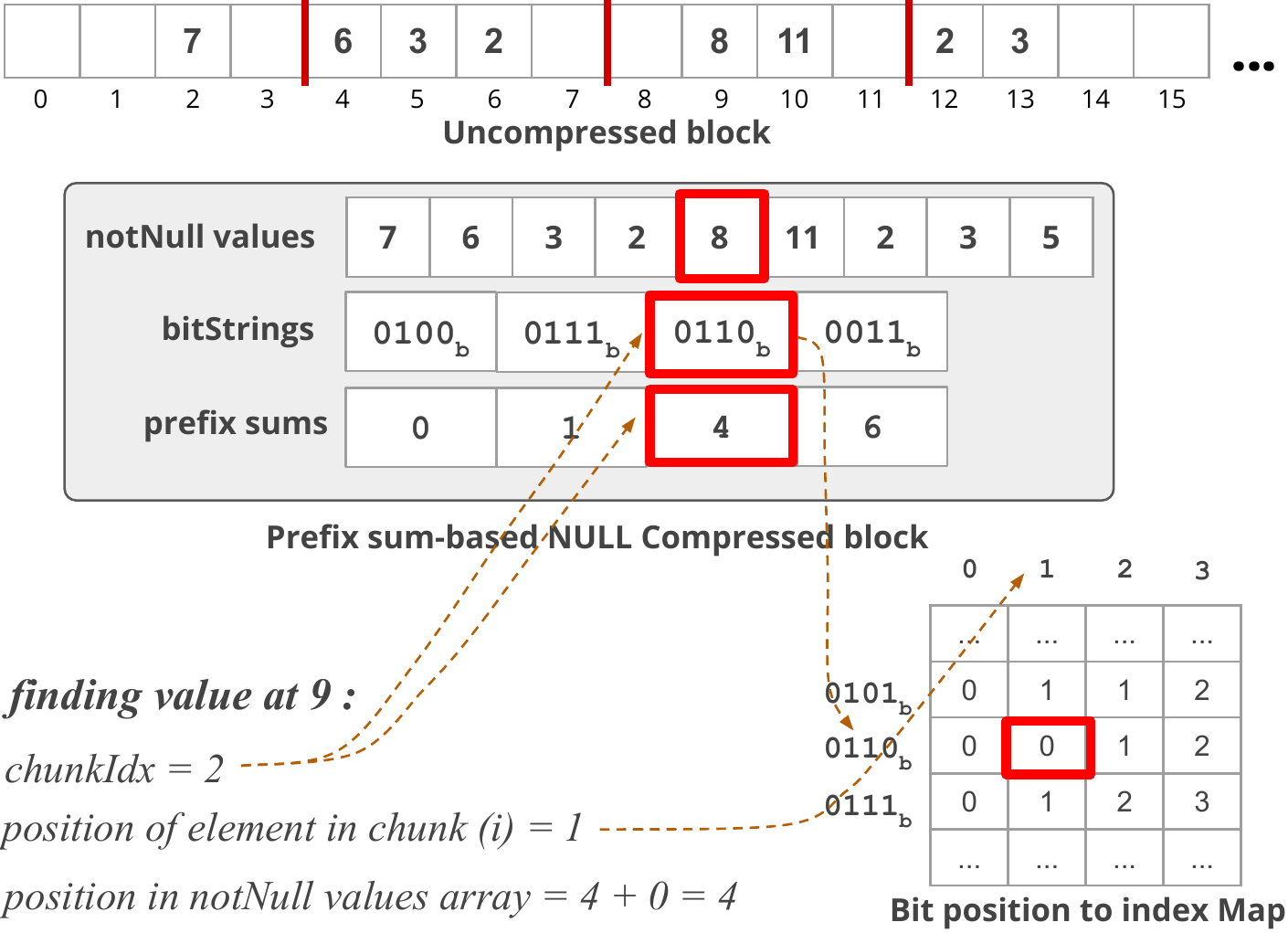}\hspace*{\fill}
	\captionsetup{justification=centering}
	\vspace{-3pt}
	\caption{NULL compression using a simplified Jacobson's bit vector rank index with chunk size 4.}
	\label{fig:null2}
	\vspace{-5pt}
\end{figure}

Abadi's third design, that uses a bit vector, already supports checking if the value at $p$ is NULL.
To support rank queries, we enhance this design with a simplified version of Jacobson's bit vector index~\cite{jacobson:bitvector, jacobson:thesis}.
Figure~\ref{fig:null2} shows this design. 
In addition to the array of non-NULL values and the bit-string, we store prefix sums 
for each $c$ (16 by default) elements in a block of a column, i.e., we divide the block 
into chunks of size $c$. 
The prefix sum holds the number of non-NULL elements before the current chunk. 
We also maintain a pre-populated static 2D \emph{bit-string-position-count map} $M$ 
with $2^c*c$ cells. $M[b, i]$ is the number of 1s before the $i$'th bit of a c-length bit string $b$.
Let $p$ be the offset which is non-NULL and $b$ the c-length bit string chunk
in the bit vector that $p$ belongs to, and $ps$ the array storing prefix sums in a block. 
Then $\text{rank(}p\text{)} = ps[p / c] + M[b, p\mod c]$.

The choice of $c$ affects how big the pre-populated map is.
A second parameter in this scheme is the number of bits $m$ used for each prefix sum value, which determines how large a block we are compressing and the overhead this scheme has for each
element. For an arbitrary $m, c$, we require: (i) $2^c*c*\lceil\log(c)/8\rceil$ byte size map, because the map has $2^c*c$ cells and needs to 
store  a $\log(c)$-bit count value in each cell; (ii) we can compress a block of size $2^m$; and (iii) we store one prefix 
sum for  each $c$ elements, so incur a cost of $m/c$ extra bits per element. 
By default we choose $m=16, c=16$. We require $2^c*c*1=1$MB-size map, can compress $2^m=64$K blocks, and incur $m/c=1$ extra bit overhead 
for each element, so increase the overhead of reference~\cite{abadi-sparse-col}'s scheme from 1 to only 2 bits per element (but provide constant time access to non-NULL values).

\section{List-based Processing}
\label{sec:list-based-processing}

We next motivate our list-based processor by 
discussing limitations of 
traditional Volcano-style tuple-at-a-time processors 
and block-based processors of columnar RDBMSs when processing n-n joins.

\begin{example}
\label{ex:proc-example}
Consider the following query. P, F, S, and O abbreviate \texttt{PERSON}, \texttt{FOLLOWS}, \texttt{STUDYAT}, and \texttt{ORGANISATION}.
{\em 
\begin{lstlisting}[numbers=none,  showstringspaces=false,belowskip=0pt ]
MATCH (a:P)$-$[:F]$\rightarrow$(b:P)$-$[:F]$\rightarrow$(c:P)$-$[:S]$\rightarrow$(d:O)
WHERE a.age > 50 and d.name = "UW"  RETURN *\end{lstlisting}
}
\end{example}

\begin{figure}
	\centering
	\includegraphics[width=1\columnwidth]{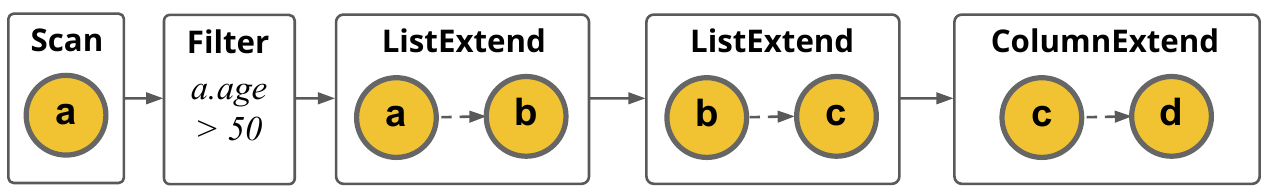}
	\vspace{-14pt}
	\caption{Query plan for the query in Example~\ref{ex:proc-example}.}
	\label{fig:proc-qp}
	\vspace{-10pt}
\end{figure}

\noindent Consider a simple plan for this query shown in Figure~\ref{fig:proc-qp}, which is
akin to a left-deep plan in an RDBMS, on a graph where \texttt{FOLLOWS} are n-n
edges and \texttt{STUDYAT} edges have single cardinality.
Volcano-style tuple-at-a-time processing~\cite{volcano}, which some GDBMSs adopt~\cite{neo4j, mhedhbi:sqs},
is efficient in terms of how much data is copied to the intermediate tuple. 
Suppose the scan matches \texttt{a} to $a_1$ and $a_1$ extends to $k_1$ many \texttt{b}'s, $b_1 \ldots b_{k_1}$, and
each $b_i$ extends to $k_2$ many \texttt{c}'s to $b_{ik_2} ..., b_{(i+1)k_2}$ (let us ignore 
the \texttt{d} extension for now). Although this generates $k_1 \times k_2$ tuples, the value $a_1$ would be copied only once to the tuple.
This is an important advantage for processing n-n joins.
However, Volcano-style processors are known to achieve low CPU and cache utility as processing 
is intermixed with many iterator calls. 

Column-oriented RDBMSs instead adopt block-based processors~\cite{boncz-phd, monet-2decades}, which process an entire block at a time in operators.
\iflongversion
\footnote{Block-based processing has been called 
vectorized processing in the original work on MonetDB/X100 from CWI~\cite{dbms-cache}.
We use the term block-based not to confuse it with SIMD vectorized instructions. 
For example, a new
column store DuckDB~\cite{duckdb} also from CWI uses block-based processing but without SIMD instructions. 
Note however that operators in block-based processors can use SIMD instructions
as they process multiple tuples at a time inside loops.}
\fi
Block sizes are fixed length, e.g. 1024 tuples~\cite{duckdb, boncz-monet-vectorized}.
While processing blocks of tuples, operators read consecutive memory locations, achieving good cache locality, and 
perform computations inside loops over arrays which is efficient on modern CPUs. 
However, traditional block-based processors have two shortcomings for GDBMSs. 
(1) For n-n joins, block-based processing requires more data copying into intermediate data
structures than tuple-at-a-time processing.
Suppose for simplicity a block size of $k_2$ and $k_1$$<$$k_2$. In our example,
the scan would output an array $a:[a_1]$, the first join would output $a:
[a_1, ..., a_1]$, $b:[b_{1}, ..., b_{k_1}]$ blocks, and the second join would output
$a:[a_1, ..., a_1]$, 
$b:[b_{1}, ...,b_{1}]$,
$c:[c_{1},...,c_{k_2}]$, 
where for example the value $a_1$ gets copied $k_2$ times
into intermediate arrays. 
(2) Traditional block-based 
processors do not exploit the list-based data organization of GDBMSs.
Specifically, the adjacency lists that are used by join operators are already stored consecutively in memory, 
which can be exploited to avoid materializing these lists into blocks.

We developed a new block-based processor called {\em list-based processor} (LBP), which
we next describe. LBP uses {\em factorized representation of intermediate tuples}~\cite{bakibayev:fdb, fct-databases, olteanu:tods} to address the data copying problem and uses block sizes set to the lengths of adjacency lists in the database, to exploit list-based data storage in GDBMSs. 

\subsection{Intermediate Tuple Set Representation}
Traditional block-based processors represent intermediate data as 
a set of {\em flat} tuples in a single group of blocks/arrays.  
In our example we had three variables \texttt{a}, \texttt{b}, and \texttt{c} corresponding to three arrays. The values 
at position $i$ of all arrays form a single tuple. Therefore to represent the tuples that are produced by n-n joins,
repetitions of values are necessary. To address these repetitions we adopt a {\em factorized tuple set representation}
scheme~\cite{olteanu:tods}. Instead of flat tuples, factorized representation systems represent tuples as 
unions of Cartesian products. For example, 
the $k_2$ flat tuples $[(a_1, b_1, c_1) \cup (a_1, b_1, c_2) \cup  ... \cup  (a_1, b_1, c_{k_2})]$
from above can be represented more succinctly in a factorized form as: $[(a_1) \times (b_1) \times(c_{1} \cup  ... \cup c_{k_2})]$. 

To adopt factorization in block-based processing, we instead use multiple groups of blocks, which
we call {\em list groups}, to represent intermediate data. Each list group has a \texttt{curIdx} field and can be in one of two states:
\begin{squishedlist}
\item Flat: If \texttt{curIdx} $\ge 0$,  
the list group is flattened and represents a single tuple that consists of the \texttt{curIdx}'th values in the blocks.

\item Unflat list of tuples: If \texttt{curIdx} $=$$-1$, the list groups represent as many tuples as the size of
the blocks it contains.
\end{squishedlist}
We call the union of list groups {\em intermediate chunk}, which represents
a set of intermediate tuples as the Cartesian product of each tuple that each list group represents.

\begin{figure}
	\centering
	\includegraphics[width=1.0\columnwidth]{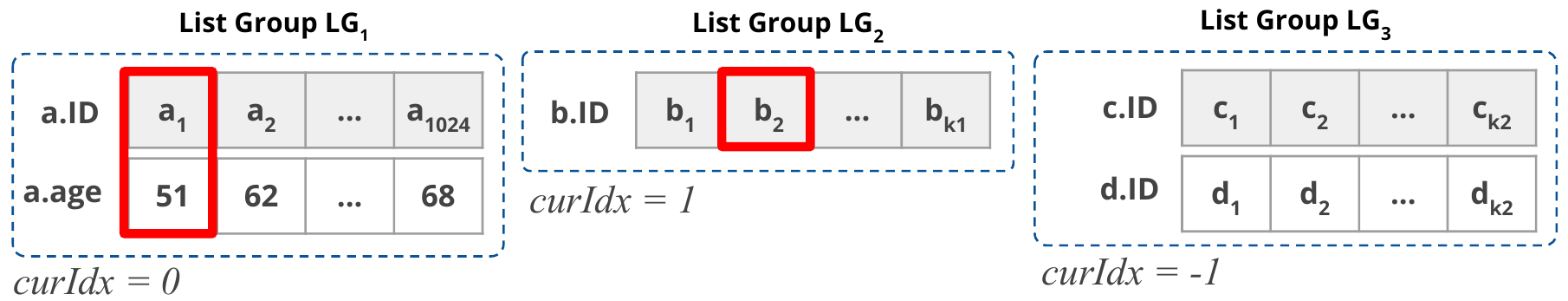}
	\vspace{-15pt}
	\captionsetup{justification=centering}
	\caption{Intermediate chunk for the query in Example~\ref{ex:proc-example}. The first two list groups are flattened to single tuples, while the last one represents $k_2$ many tuples.}
	\label{fig:list-groups}
	\vspace{-5pt}
\end{figure}

\vspace{-2pt}
\begin{example}
\label{ex:intermediate-chunk}
Figure~\ref{fig:list-groups} shows an intermediate chunk, that consists of three list groups.
The first two
groups are flattened and the last is unflat. In its current state, the intermediate chunk represents $k_2$ 
intermediate tuples as:
$(a_1, 51)  \times (b_{2}) \times ((c_{1}, d_{1}) \cup ... \cup (c_{2}, d_{2}))$.
\end{example}

In addition, instead of using fixed-length blocks as in existing block-based processors, the blocks in each group
can take different lengths, which are aligned to the lengths of adjacency lists in the database.
As we shortly explain, this allows us to avoid materializing adjacency lists into the blocks.

\subsection{Operators}
\label{subsec:operators}
We next give a description of the main relational operators we implemented to process intermediate chunks in LBP.

\noindent{{\bf \texttt{Scan}:} Scans are the ame as before and read a fixed size (1024 by default) nodeIDs into 
a block in a list group.}

\noindent{{\bf \texttt{ListExtend} and \texttt{ColumnExtend}}:} 
In contrast to a single \texttt{Join} operator that implements index nested loop join algorithm using the adjacency list indices, such as \texttt{Expand} of Neo4j, 
we have two joins. 

\texttt{ListExtend} is used to perform joins from a node, say, $a$ to nodes $b$ over 1-n or n-n edges $e$. 
The input list group $LG_a$ that holds the block of $a$ values can be flat or unflat.
If $LG_a$ is not flat, \texttt{ListExtend} first flattens it, i.e., sets the \texttt{curIdx} field of the list group to 0. It then loops through each $a$ value, say, $a_{\ell}$, and extends it to 
the set of $b$ and $e$ values using $a_{\ell}$'s adjacency list $Adj_{a_{\ell}}$. 
The blocks holding $b$ and $e$ values are put to a new list group, $LG_b$.
This allows factoring out a list of $b$ and $e$ values for a single $a$ value. 
The lengths of all blocks in $LG_b$, including those storing $b$ and $e$ as well as blocks that may be added later, will be equal to the length of $Adj_{a_{\ell}}$. This contrasts with fixed
block sizes in existing block-based processors. In addition, we exploit that
$Adj_{a_{\ell}}$ already stores $b$ and $e$ values as lists,
and do not copy these to the intermediate chunk.
Instead, the $b$ and $e$ blocks simply points to $Adj_{a_{\ell}}$.

\texttt{ColumnExtend} is used to perform 1-1 or n-1 joins. 
We call the operator \texttt{ColumnExtend} because recall from Section~\ref{subsub:vertex-columns}
that we store such edges in vanilla vertex {\em columns}. 
Suppose now that each $a$ can extend to at most one $b$ node.
\texttt{ColumnExtend} expects a block of unflat $a$ values. That is, it expects $LG_a$ to be  
unflat and adds two new blocks into $LG_a$, for storing 
$b$ and $e$, that are of the same length as $a$'s block (so unlike \texttt{ListExtend} 
does not create a new list group). 
Inside a for loop, \texttt{ColumnExtend}
copies the matching $e$ and $b$ of each $a$ from the vertex column to these two blocks. 
Note that because each $a$ value has a single $b$ and $e$ value, these values do not need to be factored out.

\noindent{\texttt{\textbf{Filter}:}}
LBP requires a more complex filter operator than those in existing block-based processors.
In particular, in traditional block based processors, binary expressions, such as a comparison expression, 
can always assume that their inputs are two blocks of values.
Instead, now binary expressions
need to operate on three possible value combinations: two flat, two lists or one list and one flat, 
because any of the two blocks can now be in a flattened list group.

\noindent{\texttt{\textbf{Group By And Aggregate:}}}
We omit a detailed description here and refer the reader to our code base~\cite{graphflowdb-github}.
Briefly, similar to \texttt{Filter}, 
\texttt{Group By And Aggregate} needs to consider whether
the values it should group by or aggregate are flat or not, and 
performs a group by and aggregation on possibly multiple factorized tuples.
Factorization allows LBP to sometimes 
perform fast group by and aggregations, similar to prior techniques that compute aggregations on compressed data~\cite{design-imp-book, abadi-col-comp}. For example, count(*) simply multiplies the sizes of each list group to 
compute the number of tuples represented by each 
intermediate chunk it receives.

\vspace{-3pt}
\begin{example}
\label{ex:operators}
Continuing our example, 
the three list groups in Figure~\ref{fig:list-groups} are an example  intermediate chunk output by the \texttt{ColumnExtend} operator in the plan from Figure~\ref{fig:proc-qp}.
In this, the initial \texttt{Scan} and \texttt{Filter} have filled the 1024-size $a$ and $a.age$ blocks in $LG_1$.
The first \texttt{ListExtend} has: (i) flattened $LG_1$ to tuple $(a_1, 51)$; and (ii) filled a 
block of $k_1$ $b$ values in a new list group $LG_2$. The second \texttt{ListExtend} has (i) flattened $LG_2$ and iterated
over it once, so its \texttt{curIdx} field is 1, and $LG_2$ now represents the tuple $(b_2)$; and (ii) has 
filled a  block of $k_2$ $c$ values in a new list group $LG_3$. Finally, the last \texttt{ColumnExtend} 
fills a block of $k_2$ $d$ values, also in $LG_3$, by extending
each $c_j$ value to one $d_j$ value through the single cardinality \texttt{STUDY_AT} edges.
\end{example}

\vspace{-3pt}
\section{Updates and Query Optimization}
\label{sec:implementation}

Although we do not focus on handling updates and query optimization within the scope 
of this paper, these components require further considerations in a complete integration of our techniques. 
As in columnar RDBMSs, the columnar storage techniques we covered are read-optimized
and necessarily add several complexities to updates~\cite{design-imp-book}. 
First recall from Section~\ref{subsubsec:csr-storage} that CSR data structure for storing adjacency list indexes are effectively sorted structures that are compressed by run-length encoding. 
So handling deletions or insertions requires resorting the CSRs and recalculating the CSR offsets. 
Insertion of edge properties in single-directional property
pages are append only and do not require any sorting. Insertions to vertex columns are also simple as these too are unsorted structure. 
However, deletions of nodes or edges, require leaving gaps 
in vertex columns and single-directional property pages. This requires keeping track of these gaps and 
reusing them for new insertions. Note
that this is also how node deletions are handled in Neo4j~\cite{neo4j-deletes}. 
Finally, the null compression scheme we adopted requires three updates upon insertion and deletions: (i) changing the bit values in the bitstrings; (ii) re-calculating prefix sum values
for prefixes after the location of update; and (iii) shifting the non-NULL elements array.
These added complexities are an inherent trade off when integrating read-optimized techniques and can be mitigated by several existing techniques, like bulk updates or keeping a write-optimized second storage
that keeps track of recent writes, which are not immediately merged.
Positional delta trees~\cite{heman:positional}
or C-Store's write-store are examples~\cite{c-store} of the latter technique. 

Two of our techniques also require additional considerations when modifying the optimizer. First, 
our use of factorized list groups changes the size of tuples that are passed between operators, as the intermediate
tuples are now compressed. When assigning costs to plans, the compressed sizes, instead of the flattened sizes
of these tuples should be considered. In addition, scans of properties that are stored in,
say forward single-directional property pages, behave differently when the properties are scanned 
in the forward direction (e.g., after a join that has used the forward adjacency lists) vs the backward 
direction. The former leads to sequential reads while the latter to random reads. 
The optimizer should assign costs based on this criterion as well.
We leave a detailed study of how to handle updates and optimize queries under our techniques to future work.

\section{Evaluations}
\label{sec:evaluations}

We integrated our columnar techniques into GraphflowDB, an in-memory GDBMS written in Java.
We refer to this version of GraphflowDB as \texttt{GF-CL} ({\bf C}olumnar {\bf L}ist-based).  
We based our work on the publicly available version here~\cite{graphflow-code}, 
which we will refer to as \texttt{GF-RV} ({\bf R}ow-oriented {\bf V}olcano).
\texttt{GF-RV} uses 8 byte vertex and edge IDs and adopts the interpreted attribute layout to store edge and vertex properties.
 \texttt{GF-RV} also partitions adjacency lists by edge labels and stores the (edge ID, neighbour ID) pairs inside a CSR.
 We present both microbenchmark experiments comparing \texttt{GF-RV} and \texttt{GF-CL} and baseline experiments
 against Neo4j, MonetDB, and Vertica  using LDBC and JOB benchmarks. 

\vspace{-4pt}
\subsection{Experimental Setup}
\label{subsec:setup}
\noindent \textbf{Hardware Setup:} For all our experiments, we use a single machine that has two Intel E5-2670 @ 2.6GHz CPUs and 512 GB of RAM. We only use one logical core. We set the maximum size of the JVM heap to 500 GB and keep JVM's default minimum size.

\noindent \textbf{Datasets:} 
Our LBP is designed to yield benefits under join queries over 1-n and n-n
relationships. Our storage compression techniques exploit some structure in the dataset and NULLs. 
These techniques are not designed for datasets that do not depict structure,
e.g., a highly heterogenous knowledge graph, such as DBPedia. 
We choose the following datasets and queries that satisfy these requirements:

\noindent {\em LDBC:} We generated the
LDBC social network data~\cite{ldbc-snb} using scale
factors 10 and 100, which we refer to as LDBC10 and LDBC100, respectively.
In LDBC, all of the edges and edge and vertex properties are structured
but several properties and edges are very sparse. 
LDBC10 contains 30M vertices and 176.6M edges while LDBC100 contains 1.77B edges and 300M vertices. Both datasets contain 8 vertex labels, 15 edge labels and 34 (29 vertex, 5 edge) properties.

\noindent {\em JOB:} We used the IMDb movie database and the JOB benchmark~\cite{leis:job}.
Although the workload has originally been created to study optimizing join order selection, 
the dataset contains several n-n, 1-n, and 1-1 relationships between entities, like
actors, movies, and companies, and structured properties, some of which contain NULLs. 
This makes it suitable to demonstrate the benefits from our storage and compression
techniques. JOB also contains join queries over n-n
relationships, making it suitable to demonstrate benefits of LBP. 
We created a property graph version of this database and workload as follows.
IMDb contains three groups of tables: (i) {\em entity tables} representing entities, such as actors (e.g., \texttt{name} table), movies, and companies; (ii) {\em relationship tables} representing n-n relationships between the entities (e.g., the \texttt{movie\-_companies} table represents relationships between movies and companies); and (iii) {\em type tables}, which denormalize the entity or relationship tables to indicate the types of entities or relationships. 
We converted each row of an entity table to a vertex. 
Let $u$ and $v$ be vertices representing, respectively, rows $r_u$ and $r_v$ from tables $T_u$ an $T_v$. We added two sets of edges between $u$ and $v$: (i) a {\em foreign key edge} from $u$ to $v$ if the primary key of row $r_u$ is a foreign key in row $r_v$; (ii) a {\em relationship edge} between $u$ to $v$ if a row $r_{\ell}$ in a relationship table $T_{\ell}$ connects row $r_u$ and $r_v$. The final dataset can be found in our codebase~\cite{graphflowdb-github}.

\noindent {\em \texttt{FLICKR} and \texttt{WIKI}:} To enhance our microbenchmarks further, we use two additional datasets
from the popular Konect graph sets~\cite{konect} covering two application domains: a Flickr social network (\texttt{FLICKR})~\cite{mislove-2008-flickr} 
and a Wikipedia hyperlink graph between articles of the German Wikipedia (\texttt{WIKI})~\cite{wiki-konect}. 
Flickr graph has 2.3M nodes and 33.1M edges while Wikipedia graph has 2.1M nodes and 86.3M edges. 
Both datasets have timestamps as edge properties.

In each experiment, we ran our queries 5 times consecutively and report the average of the last 3 runs. 
We did not observe large variances in these experiments. 
Across all of the LDBC and JOB benchmark queries we report, the median difference
between the minimum and maximum of the 3 runs we report was 1.02\% and the largest was 25\%, which was a query in which 
the maximum run was 24ms while the minimum was 19ms.

\subsection{Memory Reduction}
\label{exp:adjacency-list-exp}

We first demonstrate the memory reduction we get from the columnar storage and compression techniques we studied using LDBC100 and IMDb.
We start with \texttt{GF-RV} and integrate one additional storage optimization step-by-step 
ending with \texttt{GF-CL}:

\begin{squishedenumerate}
	
	\item \texttt{+COLS}: Stores vertex properties in vertex columns, edge properties in single-directional
	property pages, and single cardinality edges in vertex columns (instead of CSR). 
	
	\item \texttt{+NEW-IDS}: Introduces our new vertex and edge ID schemes and factors out possible ID components (recall Section~\ref{sec:storage-optimizations}).  
	
	\item \texttt{+0-SUPR}: Implements leading 0 suppression in the components of vertex and edge IDs in adjacency lists.
	
	\item \texttt{+NULL}: Implements NULL compression of empty lists and vertex properties based on Jacobson's index. 
	
\end{squishedenumerate}

\begin{table}
	\captionsetup{justification=centering}
	\caption{Memory reductions after applying one more optimization on top of the configuration on the left.}
	\vspace{-8pt} 
	\begin{subtable}{0.45\textwidth}
		\caption{LDBC100}
		\vspace{-6px}
		\centering
		\setlength{\tabcolsep}{1pt}
		\def\arraystretch{1}
		\begin{tabular}{ |c|c|c|c|c|c|c| } 
			\hline
			& \texttt{GF-RV} & \texttt{+COLS} & \texttt{+NEW-IDS} & \texttt{+0-SUPR} & \texttt{+NULL} & \texttt{GF-CL} \\ 
			\hline 
			\hline
			\multirow{2}{40pt}{Vertex Props.}& 31.40 & 19.87 & 19.87 & 19.87  & 19.28 & \textbf{-} \\ 
			& & \textbf{+1.58x} & \textbf{-} & \textbf{-} & \textbf{+1.03x} &  \textbf{1.62x} \\ 
			\hline
			\multirow{2}{40pt}{Edge Props.}& 7.92 & 2.07 & 2.07 & 2.07  & 2.05 & \textbf{-} \\ 
			& & \textbf{+3.82x} & \textbf{-} & \textbf{-} & \textbf{+1.01x} & \textbf{3.87x} \\ 
			\hline
			\multirow{2}{40pt}{F. Adj. Lists}& 31.93 & 28.95 & 20.67 & 11.41  
			& 10.78  & \textbf{-}\\ 
			& & \textbf{+1.10x} & \textbf{+1.40x} & \textbf{+1.81x} & \textbf{+1.06x}  & \textbf{2.96x}\\ 
			\hline
			\multirow{2}{40pt}{B. Adj. Lists}& 31.29 & 31.07 & 24.93 & 13.10 & 11.41  & \textbf{-}\\ 
			& & \textbf{+1.01x} & \textbf{+1.25x} & \textbf{+1.90x} & \textbf{+1.15x}  & \textbf{2.74x}\\ 
			\hline
			\multirow{2}{40pt}{Total (GB)} & 102.56 & 81.97 & 67.55 & 46.45 & 43.54  & \textbf{-}\\ 
			& & \textbf{+1.25x} & \textbf{+1.21x} & \textbf{+1.45x} & \textbf{+1.07x}  & \textbf{2.36x}\\ 
			\hline
		\end{tabular}
		\label{table:memory-ldbc}
	\end{subtable}
	\begin{subtable}{0.45\textwidth}
		\caption{IMDb}
		\vspace{-6px}
		\centering
		\setlength{\tabcolsep}{1pt}
		\def\arraystretch{1}
		\begin{tabular}{ |c|c|c|c|c|c|c| } 
			\hline
			& \texttt{GF-RV} & \texttt{+COLS} & \texttt{+NEW-IDS} & \texttt{+0-SUPR} & \texttt{+NULL} & \texttt{GF-CL} \\ 
			\hline 
			\hline
			\multirow{2}{40pt}{Vertex Props.}& 2.54 & 1.98 & 1.98 & 1.98  & 1.96 & \textbf{-} \\ 
			& & \textbf{+1.28x} & \textbf{-} & \textbf{-} & \textbf{+1.01x} &  \textbf{1.29x} \\ 
			\hline
			\multirow{2}{40pt}{Edge Props.}& 2.81 & 1.63 & 1.63 & 1.63  & 	0.90 & \textbf{-} \\ 
			& & \textbf{+1.72x} & \textbf{-} & \textbf{-} & \textbf{+1.83x} & \textbf{3.14x} \\ 
			\hline
			\multirow{2}{40pt}{F. Adj. Lists}& 1.13 & 1.02 & 0.65 & 0.41  & 0.36  & \textbf{-}\\ 
			& & \textbf{+1.10x} & \textbf{+1.57x} & \textbf{+1.57x} & \textbf{+1.15x}  & \textbf{2.96x}\\ 
			\hline
			\multirow{2}{40pt}{B. Adj. Lists}& 1.10 & 1.10 & 0.76 & 0.50 & 0.49  & \textbf{-}\\ 
			& & \textbf{+1.00x} & \textbf{+1.45x} & \textbf{+1.51x} & \textbf{+1.01x}  & \textbf{2.20x}\\ 
			\hline
			\multirow{2}{40pt}{Total (GB)} & 7.57 & 5.74 & 5.02 & 4.53 & 3.72  & \textbf{-}\\ 
			& & \textbf{+1.32x} & \textbf{+1.14x} & \textbf{+1.11x} & 	\textbf{+1.22x}  & \textbf{2.03x}\\ 
			\hline
		\end{tabular}
		\label{table:memory-imdb}
	\end{subtable}
	\vspace{-6pt}
	\label{tbl:mem1}
\end{table}

Table~\ref{table:memory-ldbc} shows how much memory each component of the system as well as the
entire system take after each optimization. 
On LDBC, we see 2.96x and 2.74x reduction for storing forward and backward adjacency lists, respectively.
We reduce memory significantly by using the new ID scheme that factors out components, such as
edge and vertex labels, and using small size integers for positional offsets.
We also see a 1.58x reduction by storing vertex properties in columns, 
which, unlike interpreted attribute layout, saves on storing the keys of the properties explicitly.
The modest memory gains in \texttt{+COLS} for storing adjacency lists is due to the fact that 8 out of 15 edge labels in LDBC SNB are single cardinality and storing them in vertex columns is cheaper than in CSRs, as we do not need CSR offsets.
We see a reduction of 3.82x when storing edge properties in single-directional property pages.
This is primarily because \texttt{GF-RV} stores a pointer for each edge, even if the edges with a particular label  have no properties. \texttt{GF-CL} stores no columns for these edges, so incurs
no overheads and avoids storing the keys of the properties explicitly.
We see modest benefits in NULL compression since empty adjacency lists are infrequent in LDBC100
and 26 of 29 vertex properties and all of the edge properties contain no NULL values. 
Overall, we obtained a reduction of 2.36x on LDBC100, reducing 
the memory cost from 102.5GB to 43.5GB.

The reductions on IMDb are shown in Table~\ref{table:memory-imdb} and are 
broadly similar to LDBC. 
For example, we see 2.96x and 2.2x reduction factors in forward and backward lists, which is 
comparable to that of LDBC. However, there are two main differences. First, we save more by compressing the edge properties using NULL compression, because 7 of 12 edge properties in IMDb have more than 50\% null values. Second, instead of a 3.82x 
reduction by storing edge properties using single directional property columns and 
single cardinality edges in vertex col\-umns (\texttt{+COLS} column of \texttt{Edge Props} row),
the factor is now 1.72x. This is because all of the edge properties in LDBC are 4-byte integers. Instead,
IMDb has primarily string edge properties (8 out of 12 of the edge properties), so these take 
more space compared to integers. Hence, the storage savings per byte of actual data is higher in case of LDBC. 
Overall,
the total reduction factor  is 2.03x reducing the memory overheads from 7.57G to 3.72G.

\vspace{-2pt}
\subsection{Single-Directional Property Pages}
\label{exp:property-pages}

We next demonstrate the query performance benefits of 
storing edge properties in single-directional property pages.
We configure GraphflowDB in two ways: (i) \texttt{EDGE COLS:} Stores edge properties in an edge column in a randomized way as edges are given random edge IDs; (ii) \texttt{PROP PAGES:} Edge properties are stored in forward-directional property pages with $k$$=$128 . 
\iflongversion
In Appendix~\ref{app:sensitivity}, we describe a sensitivity analysis to justify this default value of $k$ that we choose in our experiments
and actual system implementation. 
\fi

\begin{table}
	\centering
	\captionsetup{justification=centering}
	\caption{Runtime (in secs) of k-hop (H) queries when using property pages (PAGE$_P$) vs edge columns (COL$_E$).}
	\label{tbl:prop-pages}
	\vspace{-5pt}
	\bgroup
	\setlength{\tabcolsep}{4pt}
	\def\arraystretch{1}
	\begin{tabular}{ |c|c|c|c|c|c|c|c| } 
		\hline
		&& \multicolumn{2}{|c|}{\texttt{LDBC100}} & \multicolumn{2}{|c|}{\texttt{WIKI}} & \multicolumn{2}{|c|}{\texttt{FLICKR}} \\ 
		\hline
		&& \textbf{1H} & \textbf{2H} & \textbf{1H} & \textbf{2H} & \textbf{1H} & \textbf{2H} \\
		\hline
		\hline
		\multirow{3}{*}{P$_F$} & \texttt{COL$_E$}           & 0.55            & 65.22            & 2.97        & 42.92       & 1.88         & 888.30        \\ 
		\cline{2-8} & \multirow{2}{*}{  \texttt{PAGE$_P$}}  & 0.16            & 34.22            & 0.96        & 16.48       & 0.42         & 189.39        \\ 
		&                                               & \textbf{3.4x}   & \textbf{1.9x}    & \textbf{3.1x} & \textbf{2.6x} & \textbf{4.5x} & \textbf{4.7x}       \\
		\hline
		\multirow{3}{*}{P$_B$} & \texttt{COL$_E$}           & 1.23          & 131.01        &  6.33       & 99.28       & 2.40 		   & 1009.84       \\ 
		\cline{2-8} &\multirow{2}{*}{    \texttt{PAGE$_P$}} & 1.29          & 134.43        &  6.10       & 91.75       & 2.25         & 1183.14       \\  
		&                                       & \textbf{0.9x} & \textbf{1.0x} & \textbf{1.0x} & \textbf{1.1x} & \textbf{1.1x} & \textbf{0.9x}  \\\hline
	\end{tabular}
	\egroup
	\vspace{-5pt}
\end{table}

We use \texttt{LDBC100}, \texttt{WIKI}, and \texttt{FLICKR} datasets. As our workload, we use 1- and 2-hop queries, i.e., queries that enumerate all edges and 2-paths, with predicates on the edges. 
For LDBC, the paths enumerate \texttt{Knows} edges (\texttt{WIKI} and \texttt{FLICKR} contain only one edge label). 
1-hop query compares the edge's timestamp for \texttt{WIKI} and \texttt{FLICKR} and the \texttt{creationDate} property for LDBC to be greater than a constant. 2-hop query compares the property of each query edge to be greater than the previous edge's property. Since \texttt{WIKI} contains prohibitively many 2-hops we put a predicate on the source and destination nodes to make queries finish within reasonable time.
For each query and configurations, we consider two plans: (i) the {\em forward plan} that matches vertices from left to right in forward direction; (ii) the {\em backward plan} that matches in reverse order. 

Forward plans perform sequential reads of properties under \texttt{PROP-PAGES}, achieving good CPU cache locality. Therefore, they are expected to be more performant than backward plans under \texttt{PROP-PAGES} as well as both the plans plans under \texttt{EDGE COLS}, which all lead to random reads. We also expect backward plans to behave similarly under both configurations.
Table~\ref{tbl:prop-pages} shows our results. 
Observe that forward plans under \texttt{PROP-PAGES} is between 1.9x to 4.7x
faster than the forward plans under \texttt{EDGE COLS} and are also faster than the backward plans under \texttt{PROP-PAGES}.
In contrast, the performance of both backward plans are comparable. This is because 
neither edge columns nor forward-directional property pages provide any locality for accessing properties
in order of backward adjacency lists. This confirms our claim in Section~\ref{sec:edge-property-columns} that \texttt{PROP-PAGES}
is a better design than using vanilla edge columns.  

\vspace{-2pt}
\iflongversion
\subsection{Vertex Columns for Single Cardinality Edges}
\label{exp:single-cardinality}

In Section~\ref{exp:adjacency-list-exp}, we showed the memory gains of storing single cardinality edges in vertex columns.
This storage also improves performance because the system can directly access the edge without an indirection through a CSR.
We next demonstrate this benefit under two settings: (i) when empty lists (or edges because of single cardinality) are not NULL compressed; and (ii) when they are NULL compressed. We create 4 configurations of GraphflowDB:

\begin{squishedenumerate}
	\item \texttt{V-COL-UNC:} Single cardinality edge label edges are stored in vertex columns and are not compressed. This is equivalent to \texttt{+OMIT} configuration in Section~\ref{exp:adjacency-list-exp}.

	\item \texttt{CSR-UNC:} Single cardinality edge label edges are stored in CSR format and are not compressed.

	\item \texttt{V-COL-C:} Null compressed version of \texttt{V-COL-UNC}. This is equivalent to \texttt{+NULL} configuration in Section~\ref{exp:adjacency-list-exp}.

	\item \texttt{CSR-C:} Null compressed version of \texttt{CSR-UNC}.
\end{squishedenumerate}

We use the LDBC datasets only because the other datasets do not contain single cardinality edges. We use LDBC100. The workload consists of simple 1-, 2-, and 3-hop queries on the \texttt{replyOf} edge between \texttt{Comment} vertices.
To ensure that the joins are the dominant operation in these queries, the queries do not contain any predicates and 
return as output the count star aggregation. We evaluate each query with a plan that performs the joins in the forward direction.

\begin{table}[t!]
	\captionsetup{justification=centering}
	\caption{Vertex property columns vs. 2-level CSR adjacency lists for storing single cardinality edges: Query runtime is in seconds and memory usage in MBs.}
	\centering
	\bgroup
	\setlength{\tabcolsep}{5pt}
	\def\arraystretch{1}
		\begin{tabular}{ |p{0.08\textwidth}|c|c|c|c| }
			\hline
			& \textbf{1-hop} & \textbf{2-hop} & \textbf{3-hop} & \textbf{Mem (in MB)} \\ 
			\hline \hline
			\texttt{CSR-UNC}& 7.03 & 9.13 & 9.60 & 1266.56 \\ 
			\hline
			\multirow{2}{*}{\texttt{V-COL-UNC}}& 4.34 & 5.80 & 5.85 & 839.93 \\ 
			& \textbf{1.62x} & \textbf{1.57x} & \textbf{1.64x} & \textbf{1.51x} \\ 
			\hline
			\texttt{CSR-C}& 7.78 & 10.40 & 11.23 & 905.23 \\ 
			\hline
			\multirow{2}{*}{\texttt{V-COL-C}}& 5.23 & 8.28 & 8.41 & 478.86 \\ 
			& \textbf{1.49x} & \textbf{1.26x} & \textbf{1.34x} & \textbf{1.89x} \\ 
			\hline
		\end{tabular}
		\egroup
	\label{tbl:spp}
\end{table}

Table~\ref{tbl:spp} shows the result of queries under each system configuration. We observe up to 1.62x performance gains between uncompressed variants of vertex columns and CSR (i.e., \texttt{V-COL-UNC} vs \texttt{CSR-UNC}) and up to 1.49x gains between NULL compressed variants (i.e.,  \texttt{V-COL-C} vs \texttt{CSR-C}). 
These results verify that using vertex columns for single-cardinality edges not only saves space, {\em but also improves query performance irrespective of whether or not the edges/lists are NULL compressed or not}.
Recall that in Section~\ref{exp:adjacency-list-exp}, we had reported that NULL compression leads to modest memory reduction when
we look at the size reduction of the entire database (by 1.07x). This was because majority of the edges and 
properties are not sparse in LDBC. However, for the storage cost of specific edges, we  
observe major reductions. 
The last column of Table~\ref{tbl:spp} reports  the size of storing \texttt{replyOf} edges under each system configuration.
In LDBC100, 50.5\% of the \texttt{replyOf} forward adjacency lists are empty.
Observe that NULL compressing these lists lead to 1.75x memory reduction when using vertex columns
(from 839.93MB vs 478.86MB). We note that the reduction is lower, by 1.4x, if we use CSRs
because CSRs incur the cost of storing extra offsets which cannot be compressed if we want to maintain constant
time access to adjacency lists.
\else
\fi

\vspace{-2pt}
\subsection{Null Compression}
\label{exp:prefixSum}

We demonstrate the memory/performance trade-off of our NULL compression scheme on sparse vertex property columns.
We create multiple versions of the LDBC100, with the \texttt{creationDate} property of \texttt{Comment} 
vertices containing different percentage of NULL values. LDBC100 contains 220M \texttt{Comment} vertices,
so our column has 220M entries. We use the following 1-hop query: 
MATCH (a:\texttt{Person})$-$[e:\texttt{Likes}]$\rightarrow$(b:\texttt{Comment}) \texttt{RETURN} b.\texttt{creationDate}. This query is evaluated with a simple plan that scans a, extends to b, and then a sink operator
that reads b.\texttt{creationDate}. We compare the query performance and the memory cost of storing the 
\texttt{creationDate} column, when it is stored in three different ways: (i) \texttt{J-NULL} 
compresses the column using Jacobson's bit index with default configuration
(m=16, c=16); (ii) \texttt{Vanilla-NULL} is the vanilla bit string-based scheme from reference~\cite{abadi-sparse-col}; and (ii) \texttt{Uncompres\-sed} stores
the column in an uncompressed format.
\iflongversion
In Appendix~\ref{app:sensitivity},  
we demonstrate a sensitivity analysis for \texttt{J-NULL} running under different
m and c values. This experiment demonstrates that read performance is insensitive to these parameters. 
The memory overhead increases as $m$ increases, albeit marginally. So a reasonable choice is picking 
$m=c=16$, which incurs 1 bit extra overhead per element for storing prefix sums.
\fi

\begin{figure}
	\centering
	\includegraphics[scale=1]{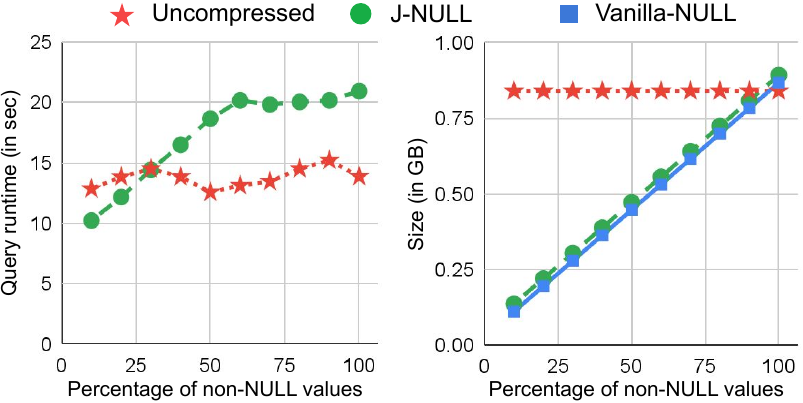}
	\captionsetup{justification=centering}
	\vspace{-6pt}
	\caption{Query performance and memory consumption  when storing a vertex property column
	as uncompressed, compressed with Jacobson's scheme, and the vanilla bit string scheme from Abadi, under different density levels.}
	\label{fig:pref-stress}
	\vspace{-10pt}
\end{figure}

Figure~\ref{fig:pref-stress} shows the memory usage and query performance under three different configurations. 
Recall that with default configuration \texttt{J-NULL} requires slightly 
more memory than \texttt{Vanilla-NULL}, 2 bits per element instead of 1 bit. 
As expected the performance of \texttt{J-NULL} is slightly slower than
\texttt{Uncompres\-sed}, between 1.19x and 1.51x, but much faster than \texttt{Vanilla-NULL}, which
was $>$20x slower than \texttt{J-NULL} and is therefore omitted in Figure~\ref{fig:pref-stress}. 
Interestingly, when the column is sparse enough (with >70\% NULL values), 
\texttt{J-NULL} can even outperform \texttt{Uncompres\-sed}.
This is because when the column is very sparse, accesses are often to NULL elements, which takes one access for reading
the bit value of the element. When the bit value is \texttt{0}, 
iterators return a global NULL value which is likely to be in the CPU cache. Instead, 
\texttt{Uncompres\-sed} always returns the value at element's cell,
which has a higher chance of a CPU cache miss. 

\subsection{List-based Processor}
\label{exp:list-based}

We next present experiments demonstrating the performance benefits of LBP
against a traditional Volcano-style tuple-at-a-time processor, which are adopted in existing systems, 
like Neo4j~\cite{neo4j} or MemGraph~\cite{memgraph}. 
LBP has three advantages over traditional tuple-at-a-time processor: 
(1) all primitive computations over data happen inside loops as in block-based operators;  
(2) the join operator can avoid copies of edge ID-neighbour ID pairs into intermediate tuples, 
exploiting the list-based storage;  and 
(3) we can perform group-by and aggregation operations directly on compressed data. 
We present two separate sets  of experiments that demonstrate the benefits from these three factors.
To ensure that our experiments only test differences due to query processing techniques, 
we integrated our columnar storage
and compression techniques into \texttt{GF-RV} (recall that this is GraphflowDB with row-based storage and 
Volcano-style processor). We  call this version \texttt{GF-CV}, for {\bf C}olumnar {\bf V}olcano.

We use LDBC100, Wikipedia, and Flickr datasets. 
In our first experiment, we take 1-, 2-, and 3-hop queries (as in Section~\ref{exp:property-pages},
we use the \texttt{Knows} edges in LDBC100), where the last
edge in the path has a predicate to be greater than a constant (e.g., e.\texttt{date} $>$ $c$).
For both \texttt{GF-CV} and \texttt{GF-CL},
we consider the standard plan that scans the left most node, extends right to match the entire
path, and a final \texttt{Filter} on the date property of the last extended edge. A major part of the work
in these plans happen at the final join and filter operation, therefore these plans allow us to measure
the performance benefits of performing computations inside loops and avoiding data copying in joins.
Our results are shown in the FILTER rows of Table~\ref{tbl:list-volcano}. 
We see that \texttt{GF-CL} outperforms \texttt{GF-CV} by large margins,
between 2.7x and 15.2x. 

In our second experiment, we demonstrate the benefits of performing fast aggregations over 
compressed intermediate results. We modify the previous queries by removing the 
predicate and instead add a return value of count(*).
We use the same plans as before except we change the last \texttt{Filter} operator 
with a \texttt{GroupBy} operator. 
Our results for aggregation are shown in the COUNT(*) rows of Table \ref{tbl:list-volcano}. Observe that the improvements 
are much more significant now, up to close to three orders of magnitude on \texttt{Wiki} (by 905.1x). The primary advantage
of \texttt{GF-CL} is now that the counting happens on compressed intermediate
results. 

\begin{table}
	\centering	
	\captionsetup{justification=centering}
	\caption{Runtime (ms) of  \texttt{GF-RV}  and \texttt{GF-CL} (LBP) plans.}
	\vspace{-5px}
	\bgroup
	\setlength{\tabcolsep}{5pt}
	\def\arraystretch{1}
	\begin{tabular}{ |c|c|c|c|c|c| }
		\hline
		&&& \textbf{1-hop}  & \textbf{2-hop} & \textbf{3-hop} \\ 
		\hline
		\multirow{6}{*}{\texttt{LDBC100}} & \multirow{3}{*}{\texttt{FILTER}} & \texttt{GF-CV} & 24.6 & 1470.5 & 40252.4 \\ 
		\cline{3-6}
		&&\multirow{2}{*}{\texttt{GF-CL}}& 7.7 & 116.2 & 2647.3 \\ 
		&&& \textbf{3.2x} & \textbf{12.7x} & \textbf{15.2x} \\ 
		\cline{2-6}
		&\multirow{3}{*}{\texttt{COUNT(*)}} & \texttt{GF-CV} & 13.4 & 241.9 & 6947.3 \\ 
		\cline{3-6}
		&&\multirow{2}{*}{\texttt{GF-CL}}& 4.2 & 18.9 & 357.9 \\ 
		&&& \textbf{3.2x} & \textbf{12.8x} & \textbf{19.4x} \\ 
		\hline
		\multirow{6}{*}{\texttt{FLICKR}} & \multirow{3}{*}{\texttt{FILTER}} & \texttt{GF-CV} & 32.6 & 1300.0 & 14864.0 \\ 
		\cline{3-6}
		&&\multirow{2}{*}{\texttt{GF-CL}}& 12.2 & 95.3 & 1194.7 \\ 
		&&& \textbf{2.7x} & \textbf{13.7x} & \textbf{12.4x} \\ 
		\cline{2-6}
		&\multirow{3}{*}{\texttt{COUNT(*)}} & \texttt{GF-CV} & 35.3 & 519.2 & 4162.5\\ 
		\cline{3-6}
		&&\multirow{2}{*}{\texttt{GF-CL}}& 16.9 & 23.4 & 51.7 \\ 
		&&& \textbf{2.1x} & \textbf{21.4x} & \textbf{80.6x} \\ 
		\hline
		\multirow{6}{*}{\texttt{WIKI}} & \multirow{3}{*}{\texttt{FILTER}} & \texttt{GF-CV} & 35.8 & 4500.2 & 236930.2 \\ 
		\cline{3-6}
		&&\multirow{2}{*}{\texttt{GF-CL}}& 11.9 & 1192.5 & 20329.3 \\ 
		&&& \textbf{2.9x} & \textbf{3.8x} & \textbf{11.7x} \\ 
		\cline{2-6}
		&\multirow{3}{*}{\texttt{COUNT(*)}} & \texttt{GF-CV} & 32.7 & 1745.2 & 109000.2 \\ 
		\cline{3-6}
		&&\multirow{2}{*}{\texttt{GF-CL}}& 19.0 & 27.6 & 120.4 \\ 
		&&& \textbf{1.7x} & \textbf{63.2x} & \textbf{905.1x} \\ 
		\hline
	\end{tabular}
	\egroup
	\label{tbl:list-volcano}
	\vspace{-10pt}
\end{table}

\subsection{Baseline System Comparisons}
\label{subsec:end-to-end}

In our final experiment, we compare the query performance of \texttt{GF-CL}
against \texttt{GF-RV}, Neo4j, which is another 
row-oriented and Volcano style GDBMSs,
and two columnar RDBMSs, MonetDB and Vertica, 
which are not tailored for n-n joins.
Our primary goal is to verify that \texttt{GF-CL} is faster than \texttt{GF-RV} also on an independent end-to-end benchmark.
We also aim to verify that \texttt{GF-RV}, on which we base our work, is already competitive with or
outperforms other baseline systems on workloads containing n-n joins. 
We used the SNB on LDBC10 and JOB, both
of which contain n-n join queries.

We used the community version v4.2 of Neo4j GDBMS~\cite{neo4j}, the community version 10.0 of Vertica~\cite{vertica-docs} and MonetDB 5 server 11.37.11~\cite{monet-source}. We note that our experiments should not be interpreted as one system being more efficient than another. It is difficult to meaningfully compare completely separate systems, e.g., all baseline systems have many tunable parameters, and some have more efficient enterprise versions. 
For all baseline systems, we map their storage to an in-memory filesystem, set number of CPUs to 1 and disable spilling intermediate files to disk.
We maintain 2 copies of edge tables for Vertica and MonetDB, sorted by the source and destination vertexIDs, respectively. 
For \texttt{GF-RV} and \texttt{GF-CL}, we use the best left-deep plan we could manually pick, which was obvious in most cases. For example, LDBC path queries start from a particular vertex ID, so the 
best join orders start from that vertex and iteratively extend in the same direction. 
For Vertica, MonetDB, and Neo4j, we use the better of the systems' default plans and the left-deep that is 
equivalent to the one we use in \texttt{GF-RV} and \texttt{GF-CL}.

\subsubsection{LDBC}
We use the LDBC10 dataset. GraphflowDB is a prototype system that implements
parts of the Cypher language relevant to our research, so
lack several features that LDBC queries exercise. The system currently 
has support for select-project-join queries and a limited form of aggregations,
where joins are expressed as fixed-length subgraph patterns in the MATCH clause. We modified the 
Interactive Complex Reads (IC) and Interactive Short Reads (IS) queries from LDBC~\cite{ldbc-snb}
in order to be able to run them.
Specifically GraphflowDB does not support variable length queries that
search for joins between a minimum and maximum length, which we set to the maximum length to make them fixed-length instead, and shortest path queries, which we removed from the benchmark. We also removed predicates that 
check  the existence or non-existence of edges between nodes and the ORDER BY clauses.  
\iflongversion
Our exact queries are given in Appendix~\ref{app:ldbc-snb}.
\fi

\iflongversion
\begin{table*}
	\vspace{-40pt}
	\begin{subtable}{1\textwidth}
		\setlength{\tabcolsep}{3.5pt}
		\def\arraystretch{1.1}
		\centering
		\begin{tabular}{ |c|c|c|c|c|c|c|c| }
			\hline
			& \textbf{IS01} & \textbf{IS02} & \textbf{IS03} & 	\textbf{IS04} & \textbf{IS05} & \textbf{IS06} & \textbf{IS07}\\ 
			\hline
				
			{\texttt{GF-CL}} & 2.7 & 3.0 & 2.2 & 36.9 & 40.6 & 69.3 & 38.3 \\ 
			\hline	
		
			\multirow{2}{*}{\texttt{GF-RV}}& 2.2 & 3.9 & 3.9 & 307.3 & 236.6 & 423.0 & 307.9 \\ 
			& \textbf{0.8x} & \textbf{1.3x} & \textbf{1.8x} & \textbf{8.3} & \textbf{5.8x} & \textbf{6.1x} & \textbf{8.0x} \\ 
			\hline

			\multirow{2}{*}{\texttt{VERTICA}} & 6.1 & 16728.2 & 7.0 & 1.5 & 45.2 & 259.2 & 24818.9 \\ 
			& \textbf{2.2x} & \textbf{5574.2x} & \textbf{3.2x} & \textbf{0.04x} & \textbf{1.1x} & \textbf{3.7x} & \textbf{647.7x} \\ 
			\hline
		
			\multirow{2}{*}{\texttt{MONET}} & 112.3 & 282.2 & 8.3 & 84.1 & 516.4 & 323.0 & 206.3 \\ 
			& \textbf{40.9x} & \textbf{94.0x} & \textbf{3.8x} & \textbf{2.3x} & \textbf{12.7x} & \textbf{4.7x} & \textbf{5.4x} \\ 
			\hline
		
			\multirow{2}{*}{\texttt{NEO4J}} & 103.1 & 117.4 & 86.1 & 12418.9 & 11665.9 & 67390.3 & 12095.2 \\ 
			& \textbf{37.5x} & \textbf{39.1x} & \textbf{39.1} & \textbf{336.6} & \textbf{287.4} & \textbf{972.3} & \textbf{315.6} \\ 
			\hline
		\end{tabular}
		\caption{LDBC IS Queries}
		\label{table:ldbc-is-individual}
	\end{subtable}\vspace{0.1 cm}
	\begin{subtable}{1\textwidth}
		\setlength{\tabcolsep}{3.5pt}
		\def\arraystretch{1.1}
		\centering
		\begin{tabular}{ |c|c|c|c|c|c|c|c|c|c|c|c| }
			\hline
			& \textbf{IC01} & \textbf{IC02} & \textbf{IC03} & \textbf{IC04} & \textbf{IC05} & \textbf{IC06} & \textbf{IC07} & \textbf{IC08} & \textbf{IC09} & \textbf{IC011} & \textbf{IC012} \\ 
			\hline
						
			{\texttt{GF-CL}} & 36.7 & 32.4 & 409.4 & 13.1 & 1565.2 & 113.0 & 3.0 & 2.6 & 1519.8 & 11.1 & 34.2 \\ 
			\hline
		
			\multirow{2}{*}{\texttt{GF-RV}}& 88.4 & 45.2 & 1521.8 & 57.3 & 8925.0 & 333.1 & 6.3 & 7.0 & 2098.1 & 19.2 & 84.9 \\ 
			& \textbf{2.4x} & \textbf{1.4x} & \textbf{3.7} & \textbf{4.4x} & \textbf{5.7x} & \textbf{3.0x} & \textbf{2.1x} & \textbf{2.7x} & \textbf{1.4x} & \textbf{1.7x} & \textbf{2.5x} \\ 
			\hline
			\multirow{2}{*}{\texttt{VERTICA}} & 257.2 & 3063.8 & 18610.3 & 1711.6 & 59351.0 & 4715.7 & 4092.2 & 2837.2 & 17276.2 & 672.9 & 5028.1 \\ 
			& \textbf{7.0x} & \textbf{94.5x} & \textbf{45.5x} & \textbf{130.5x} & \textbf{37.9x} & \textbf{41.7x} & \textbf{1348.8x} & \textbf{1094.2x} & \textbf{11.4x} & \textbf{60.9x} & \textbf{147.1x} \\ 
			\hline
		
			\multirow{2}{*}{\texttt{MONET}} & 160.3 & 323.2 & 187330.9 & 13955.1 & 165273.0 & 2783.1 & 206.3 & 920.5 & 121943.2 & 572.0 & 3251.9 \\ 
			& \textbf{4.4x} & \textbf{10.0x} & \textbf{457.6x} & \textbf{1064.3x} & \textbf{105.6x} & \textbf{24.6x} & \textbf{68.0x} & \textbf{354.8x} & \textbf{80.0x} & \textbf{51.7x} & \textbf{95.1x} \\ 
			\hline
		
			\multirow{2}{*}{\texttt{NEO4J}} & 669.3 & 170722.9 & 86231.9 & 75254.9 & \textit{TLE} & 515.4 & 95.6 & 108.3 & 219425.5 & 2804.1 & 34043.0 \\ 
			& \textbf{18.3x} & \textbf{5264.2x} & \textbf{210.6x} & \textbf{5739.4x} & \textbf{-} & \textbf{4.6x} & \textbf{31.5x} & \textbf{41.8x} & \textbf{144.4x} & \textbf{253.6x} & \textbf{996.0x} \\ 
			\hline
		\end{tabular}
		\caption{LDBC IC Queries}
		\label{table:ldbc-ic-individual}
	\end{subtable}\vspace{0.1 cm}
	\begin{subtable}{1\textwidth}
		\setlength{\tabcolsep}{3.5pt}
		\def\arraystretch{1.1}
		\centering
		\begin{tabular}{ |c|c|c|c|c|c|c|c|c|c|c|c| }
			\hline
			& \textbf{1.a} & \textbf{2.a} & \textbf{3.a} & \textbf{4.a} & \textbf{5.a} & \textbf{6.a} & \textbf{7.a} & \textbf{8.a} & \textbf{9.a} & \textbf{10.a} & \textbf{11.a} \\ 
			\hline
		
			{\texttt{GF-CL}} & 13.5 & 59.7 & 41.4 & 17.4 & 124.8 & 9.8 & 53.4 & 298.3 & 209.7 & 51.3 & 23.6 \\ 
			\hline	
		
			\multirow{2}{*}{\texttt{GF-RV}}& 50.3 & 120.4 & 38.0 & 29.4 & 460.9 & 91.4 & 77.5 & 1245.6 & 560.2 & 110.4 & 487.3   \\ 
			& \textbf{3.7x} & \textbf{2.0x} & \textbf{0.9x} & \textbf{1.7x} & \textbf{3.7x} & \textbf{9.3x} & \textbf{1.5x} & \textbf{4.2x} & \textbf{2.7x} & \textbf{2.2x} & \textbf{2.1x} \\ 
			\hline
		
			\multirow{2}{*}{\texttt{VERTICA}} & 214.8 & 329.8 & 3928.3 & 798.1 & 2591.4 & 495.2 & 72.8 & 1166.2 & 442.5 & 395.4 & 136.9 \\ 
			& \textbf{15.9x} & \textbf{5.5x} & \textbf{94.9x} & \textbf{45.9x} & \textbf{20.8x} & \textbf{50.6x} & \textbf{1.4x} & \textbf{3.9x} & \textbf{2.1x} & \textbf{7.7x} & \textbf{5.8x} \\ 
			\hline
		
			\multirow{2}{*}{\texttt{MONET}} & 36.2 & 33.0 & 36.2 & 120.2 & 232.5 & 1428.1 & 133.6 & 112.5 & 282.5 & 304.1 & 92.8 \\ 
			& \textbf{2.7x} & \textbf{0.6x} & \textbf{0.9x} & \textbf{1.3x} & \textbf{1.9x} & \textbf{145.9x} & \textbf{2.5x} & \textbf{0.4x} & \textbf{1.4x} & \textbf{5.9x} & \textbf{3.9x} \\ 
			\hline
		
			\multirow{2}{*}{\texttt{NEO4J}} & 3077.7 & 895.3 & 774.3 & 203.4 & 10727.2 & 206.7 & 6497.6 & 3451.7 & 14946.4 & 1480.7 & 2332.6 \\ 
			& \textbf{227.5x} & \textbf{15.0x} & \textbf{18.7x} & \textbf{11.7} & \textbf{85.9x} & \textbf{21.1x} & \textbf{121.8x} & \textbf{11.6x} & \textbf{71.3x} & \textbf{28.9x} & \textbf{98.8x} \\ 
			\hline
		\end{tabular}\vspace{0.1 cm}
		\begin{tabular}{ |c|c|c|c|c|c|c|c|c|c|c|c| }
			\hline
			& \textbf{12.a} & \textbf{13.a} & \textbf{14.a} & \textbf{15.a} & \textbf{16.a} & \textbf{17.a} & \textbf{18.a} & \textbf{19.a} & \textbf{20.a} & \textbf{21.a} & \textbf{22.a} \\ 
			\hline
	
			{\texttt{GF-CL}} & 58.3 & 70.0 & 14.6 & 362.5 & 15.0 & 268.5 & 548.1 & 207.8 & 12.8 & 13.2 & 28.6 \\ 
			\hline	
	
			\multirow{2}{*}{\texttt{GF-RV}} & 253.7 & 406.9 & 33.3 & 6772.3 & 34.0 & 594.6 & 1700.9 & 983.0 & 208.5 & 22.6 & 64.4 \\ 
			& \textbf{4.4x} & \textbf{5.8x} & \textbf{2.3x} & \textbf{18.7x} & \textbf{2.3x} & \textbf{2.2x} & \textbf{3.1x} & \textbf{4.7x} & \textbf{1.6x} & \textbf{1.7x} & \textbf{2.3x} \\ 
			\hline
	
			\multirow{2}{*}{\texttt{VERTICA}} & 870.2 & 286.3 & 28.2 & 2100.5 & 1028.8 & 2538.5 & 1686.0 & 4777.2 & 982.5 & 34.0 & 99.0 \\ 
			& \textbf{2.4x} & \textbf{4.1x} & \textbf{1.9x} & \textbf{5.8x} & \textbf{68.5x} & \textbf{9.5x} & \textbf{3.1x} & \textbf{23.0x} & \textbf{76.7x} & \textbf{2.6x} & \textbf{3.5x} \\ 
			\hline
	
			\multirow{2}{*}{\texttt{MONET}} & 56.7 & 1148.2 & 83.4 & 172.0 & 224.5 & 1304.3 & 868.2 & 644.0 & 7552.3 & 60.7 & 140.2 \\ 
			& \textbf{1.0x} & \textbf{16.4x} & \textbf{5.7x} & \textbf{0.5x} & \textbf{14.9x} & \textbf{4.9x} & \textbf{1.6x} & \textbf{3.1x} & \textbf{590.0x} & \textbf{4.6x} & \textbf{4.9x} \\ 
			\hline
	
			\multirow{2}{*}{\texttt{NEO4J}} & 5079.1 & 93.8 & 291.9 & 2437.4 & 4526.6 & 167.6 & 1414.8 & 12047.2 & 1849.0 & 272.4 & 317.8 \\ 
			& \textbf{87.2x} & \textbf{1.3x} & \textbf{20.1x} & \textbf{6.7x} & \textbf{301.2x} & \textbf{0.6x} & \textbf{2.6x} & \textbf{58.0x} & \textbf{144.5x} & \textbf{20.7x} & \textbf{11.1x} \\ 
			\hline
		\end{tabular}\vspace{0.1 cm}
		\begin{tabular}{ |c|c|c|c|c|c|c|c|c|c|c|c| }
			\hline
			& \textbf{23.a} & \textbf{24.a} & \textbf{25.a} & \textbf{26.a} & \textbf{27.a} & \textbf{28.a} & \textbf{29.a} & \textbf{30.a} & \textbf{31.a} & \textbf{32.a} & \textbf{33.a} \\ 
			\hline
		
			{\texttt{GF-CL}} & 14.5 & 10.8 & 107.7 & 10.5 & 10.3 & 26.1 & 5.6 & 18.3 & 112.5 & 10.0 & 52.3 \\ 
			\hline
		
			\multirow{2}{*}{\texttt{GF-RV}} & 407.8 & 47.9 & 1527.8 & 19.9 & 125.3 & 56.7 & 18.5 & 52.7 & 775.9 & 24.1 & 201.6 \\ 
			& \textbf{28.8x} & \textbf{4.4x} & \textbf{14.2x} & \textbf{1.9x} & \textbf{12.2x} & \textbf{2.2x} & \textbf{3.1} & \textbf{2.9x} & \textbf{6.9x} & \textbf{2.4x} & \textbf{3.9x} \\ 
			\hline
		
			\multirow{2}{*}{\texttt{VERTICA}} & 698.5 & 518.1 & 496.2 & 1239.9 & 231.0 & 197.3 & 3153.1 & 152.6 & 2696.3 & 193.6 & 125.3 \\ 
			& \textbf{49.4x} & \textbf{47.8x} & \textbf{4.6x} & \textbf{118.7x} & \textbf{22.5x} & \textbf{7.6x} & \textbf{529.9x} & \textbf{8.5x} & \textbf{23.9x} & \textbf{19.3x} & \textbf{2.4x} \\ 
			\hline
		
			\multirow{2}{*}{\texttt{MONET}} & 124.2 & 993.9 & 784.8 & 1736.1 & 75.9 & 323.8 & 1012.3 & 1940.2 & 848.1 & 87.7 & 88.1 \\ 
			& \textbf{8.8x} & \textbf{91.8x} & \textbf{7.3x} & \textbf{166.1x} & \textbf{7.4x} & \textbf{12.4x} & \textbf{170.1x} & \textbf{107.5x} & \textbf{7.5x} & \textbf{8.8x} & \textbf{1.7x} \\ 
			\hline
		
			\multirow{2}{*}{\texttt{NEO4J}} & 2497.1 & 3505.4 & 108.6 & 694.1 & 1276.7 & 1573.7 & 648.2 & 326.1 & 152.7 & 364.1 & 2723.8 \\ 
			& \textbf{176.5x} & \textbf{323.7x} & \textbf{1.0x} & \textbf{66.3x} & \textbf{124.4x} & \textbf{60.3x} & \textbf{108.9x} & \textbf{18.1x} & \textbf{1.4x} & \textbf{36.4x} & \textbf{52.0x} \\ 
			\hline
		\end{tabular}
		\label{table:job-individual}
		\caption{JOB Benchmark}
	\end{subtable}
	\captionsetup{justification=centering}
	\caption{Runtime in ms for running the LDBC Interactive Complex Reads (IC) and Interactive Short Reads (IS) queries and JOB Benchmark on 5 systems: (i) \texttt{GF-CL}  
	(ii) \texttt{GF-RV} 
	(iii) \texttt{VERTICA} (iv) \texttt{MONET} and (v) \texttt{NEO4J}. }
	\label{tbl:sys-comp}
\end{table*}
\fi

\begin{figure}
	\centering
	\includegraphics[width=0.95\columnwidth]{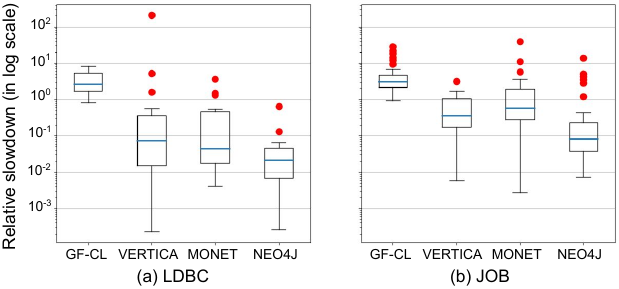}
		\captionsetup{justification=centering}
	\vspace{-5pt}
	\caption{Relative speedup/slowdown of the different systems in comparison to \texttt{GF-RV} on LDBC10. The boxplots show the 5th, 25th, 50th, 75th, and 95th percentiles.}
	\label{fig:end-to-end}
	\vspace{-13pt}
\end{figure}

Figure~\ref{fig:end-to-end}a shows the relative speedup/slowdown of the different systems in comparison to \texttt{GF-RV}.
\iflongversion
Tables~\ref{table:ldbc-is-individual} and~\ref{table:ldbc-ic-individual}, show the individual runtime numbers of each IS and IC query, respectively.
\fi 
As expected, \texttt{GF-CL} is broadly more performant than \texttt{GF-RV} on LDBC
with a median query improvement factor of 2.6x. With the exception of one query, which slows down a bit, the performance of {\em every query} improves between 1.3x to 8.3x. The improvements come from several optimizations but primarily from LBP and our columnar storage. In \texttt{GF-RV}, scanning properties requires checking equality on property keys, which are avoided in columnar storage, so we observed large improvements on queries that produce large intermediate results and perform filters, such as IC05. IC05 has 4 n-n joins starting from a node and extending in the forward direction 
and a predicate on the edges of the third join. 
\texttt{GF-CL} has several advantages that become visible here. First, \texttt{GF-CL}'s 
LBP, unlike \texttt{GF-RV}, does not copy any edge and neighbour IDs to intermediate tuples. More importantly,
LBP performs filters inside loops and \texttt{GF-CL}'s single-indexed property pages provides faster access to the edge properties that are used in the filter than \texttt{GF-RV}'s row-oriented format.
On this query, \texttt{GF-RV} takes 8.9s while \texttt{GF-CL} takes 1.6s.

As we expected, we also found other baseline systems to not be as performant as \texttt{GF-CL} or \texttt{GF-RV}.
In particular, Vertica, MonetDB, and Neo4j have median slowdown factors of 13.1x, 22.8x, and 46.1x compared to \texttt{GF-RV}. 
Although Neo4j performed slightly worse than other baselines, we also observed that there were some queries in which it outperformed Vertica and MonetDB (but not \texttt{GF-RV} or \texttt{GF-CL}) 
by a large margin. These were queries
that started from a single node, had several n-n joins, but did not generate large intermediate results,
like IS02 or IC06. On such queries, GDBMSs, both GraphflowDB and Neo4j, have the advantage
of using join operators that use the adjacency list indices to extend a set of partial matches. This 
can be highly efficient if the partial matches that are extended are small in number.
For example, the first join of IC06 extends a single \texttt{Person} node, say $p_i$, to its two-degree friends. In SQL,
this is implemented as joining a \texttt{Person} table with a \texttt{Knows} table with a predicate on the Person table to
select $p_i$. In Vertica or MonetDB, this join is performed using merge or hash joins, which requires scanning
both \texttt{Person} and \texttt{Knows} tables. Instead, Neo4j and GraphflowDB only scan the \texttt{Person}
table to find $p_i$ and then extend $p_i$ to its neighbours, without scanning all \texttt{Knows} edges.
For this, \texttt{GF-RV}, \texttt{GF-CL}, and Neo4j take 333ms, 113ms, and 515ms, while Vertica and MonetDB take 4.7s and 2.7s, respectively. We also found that all baseline systems, including Neo4j, degrade in performance
on queries with many n-n joins that generate large intermediate results. For example, on IC05 that we reviewed before, Vertica take 1 minute, MonetDB 3.25 minutes, while Neo4j took over 10 minutes. 

\vspace{-3pt}
\subsubsection{JOB}
JOB queries come in four variants and we used their first variant. We converted the JOB queries to
their Cypher equivalent following our conversion of the dataset.
Many of the JOB queries returned aggregations on strings, such as \texttt{min(name)}, where name is a string column. Since Graphflow supports
aggregations only on numeric types, we removed these aggregations. 
\iflongversion
Our final queries are in Appendix~\ref{app:job-benchmark-queries}.
\fi

Figure~\ref{fig:end-to-end}b shows the relative performance of different systems in comparison to \texttt{GF-RV}.
\iflongversion
Table~\ref{table:ldbc-ic-individual} shows individual runtime numbers of each query.
\fi 
Similar to our LDBC results, we see GF-CL to improve the performance, now by 3.1x. Again similar
to LDBC, with the exception of one query, we see consistent speed ups across all queries between 1.5x and 28.8x. Different 
from LDBC, we also see queries on which the improvement factors are much larger, i.e, >20x. In LDBC, the largest improvement factor was 8.3x. This is expected as most of the queries in JOB perform star 
joins while LDBC queries contained path queries that start from a node with a selective filter. On path queries,
our plans start from a single node and extend in one direction, in which case only the last extension can truly be factorized, so
be in unflat form. This is because each \texttt{ListExtend} that we use first flattens the previously extended node. Whereas on star queries, multiple extensions from the center node can remain unflattened. Therefore GF-CL's plans can benefit more from LBP as they can compress their intermediate tuples more.
We also see that similar to LDBC, GF-RV is more performant than the columnar RDBMSs. However, these systems are now more competitive. We noticed that one reason for this is that on star queries,
these systems's default plans are often bushy plans (27 out of 33 for MonetDB and 26 out of 33 for Vertica), 
which produce fewer intermediate tuples than GF-RV, which does not
benefit from factorization and uses left-deep plans. 
So these systems now benefit from
bushy plans which they did not in LDBC.
In contrast, on LDBC, these systems would also primarily use 
left-deep plans (only 2 out of 18 for MonetDB and 4 out of 18 for Vertica were bushy) because on these path queries, it is better to start from a single highly filtered 
node table and join iteratively in a left-deep plan to match the entire path. Finally, similar to LDBC, Neo4j is again least competitive of these baselines.

\section{Related Work}
\label{sec:rw}

Column stores~\cite{monet-2decades, c-store, boncz-vectorwise, boncz-vectorwise1} are designed primarily for OLAP queries that perform aggregations over large amounts of data. 
Work on them introduced a set of storage and query processing techniques which include use of positional offsets, schemes for compression, block-based query processing, late materialization and operations on compressed data, among others. A detailed survey of these techniques can be found in reference~\cite{design-imp-book}. This paper aims to integrate some of these techniques into in-memory GDBMSs. 

Existing GDBMSs and RDF systems usually store the graph topology in a columnar structure. This is done either by using a variant of adjacency list or CSR. Instead, systems often use row-oriented structures
to store properties, such as an interpreted attribute layout~\cite{beckmann:sparse}. For example, Neo4j~\cite{neo4j} represents the graph topology in adjacency lists that are partitioned by edge labels and stored in linked-lists, where each edge record points to the next. 
Properties of each vertex/edge are stored in a linked-list, where each property record points to the next and encodes the key, data type, and value of the property. JanusGraph too \cite{janusgraph} stores edges in adjacency lists partitioned by edge labels and properties as consecutive key-value pairs (a row-oriented format). 
These native GDBMSs adopt Volcano-style processors. In contrast, our design adopts columnar structures for vertex and edge properties and a block-based processor. 
In addition, we compress edge and vertex IDs and NULLs.

There are also several GDBMSs that are developed directly on top of an RDBMS or another database system~\cite{oracle}, such as IBM Db2 Graph~\cite{tian:ibm-db2-graph-long}, Oracle Spatial and Graph~\cite{oracle} and SAP's graph data\-base~\cite{hana-graph}.
These systems can benefit from the columnar techniques in the underlying RDBMS, which are however not optimized for graph storage and queries. For example, SAP's graph engine uses SAP HANA's columnar-storage for edge tables but these tables do not have CSR-like structures for storing edges. 

GQ-Fast~\cite{lin:gq-fast} implements a limited SQL called relationship queries that support joins of tables similar to path queries, followed with aggregations. The system stores n-n relationship in tables with CSR-like indices and heavy-weight compression of lists and has a fully pipelined query processor that uses query compilation. 
Therefore, GQ-Fast studies how some techniques in GDBMSs, specifically joins using adjacency lists, can be done in RDBMS. In contrast, we focus on studying how some techniques from 
columnar RDBMSs can be integrated into GDBMSs. 
We intended to but could not compare against GQ-Fast because the system supports a very limited set of queries (e.g., none of the LDBC queries are supported).

\iflongversion
ZipG~\cite{zipg} is a distributed compressed storage engine for property graphs that can answer queries to retrieve adjacencies as well as vertex and edge properties. ZipG is based on a compressed data structure called Succinct~\cite{succinct}. Succinct stores semi-structured data that is encoded as a set of key and list of values. For example, a vertex $v$'s properties can be stored with the $v$'s ID as the key and a list of values, corresponding to each property. 
Succinct compresses these files using suffix arrays and several secondary indices.
Although the authors report good compression rates, access to a particular record is not constant time and requires accessing secondary indexes followed by a binary search, which is 
slower than our structures.
\else
\fi

Several RDF systems also use columnar structures to store RDF data. Reference \cite{rdf-vertical} uses a set of columns, where each column store is a set of (subject, object) pairs for a unique predicate. 
However, this storage is not as optimized as the storage in GDBMSs, e.g., the edges between entities are not stored in native CSR or adjacency list format. Hexastore \cite{hexastore} improves on the idea of predicate partitioning by having a column for each RDF element (subject, predicate or object) and sorting it in 2 possible ways in B+ trees. This is similar but not as efficient as double indexing of adjacency lists in GDBMSs. RDF-3X \cite{rdf-3x} is an RDF system that stores a large triple table that is indexed in 6 B+ tree indexes over each column. Similarly, this storage is not as optimized as the native graph storages found in GDBMSs. 
Similar to our Guideline~\ref{gdln:graph-schema},
reference ~\cite{neumann:cs} also observes that graphs have structure, and certain predicates in 
RDF databases co-exist together in a node. This is similar to the property co-occurrence structure we exploit, and is
exploited in the RDF 3-X system for better cardinality estimation.

Several novel storage techniques for storing graphs are optimized for 
write-heavy workloads, such as streaming. These works propose 
data structures that try to achieve the sequential read capabilities of CSR while being write-optimized.
Examples of this include LiveGraph~\cite{zhu:live-graph}, 
Aspen~\cite{dhulipala:aspen}, and LLAMA~\cite{macko:llama}.
We focus on a read-optimized system setting and use CSR to store the graph topology
but these techniques are complementary to our work. 

Our list groups represent intermediate results in a factorized form.
Prior work on factorized representations in RDBMSs, specifically FDB~\cite{bakibayev:aggregation, bakibayev:fdb}, represents intermediate data as tries, and 
have operators that transform tries into other tries. Unlike traditional processors,
processing is not pipelined and all intermediate results are materialized. 
Instead, operators in LBP are variants of traditional block-based operators and perform computations in a pipelined fashion on batches of lists/arrays of data. 
This paper focuses on integration of columnar storage and query processing techniques into GDBMSs and does not studies how to integrate more advanced factorized processing techniques inside GDBMS. 

\section{Conclusions}
\label{sec:conclusions}

Columnar RDBMSs are read-optimized analytical systems that have introduced several storage and query processing techniques to improve the scalability and performances of RDBMSs. We studied the integration of such techniques into GDBMSs, which are also read-optimized analytical systems. 
While some techniques can be directly applied to GDBMSs, adaptation of others can be significantly
sub-optimal in terms of space and performance. 
In this paper, we first outlined a set of guidelines and desiderata for designing the storage layer 
and query processor of GDBMSs, based on the typical access patterns in GDBMSs which are significantly
different than the typical workloads of columnar RDBMSs. We then 
presented our design of columnar storage, compression, and query processing techniques that are optimized 
for in-memory GDBMSs. Specifically, we introduced a novel list-based query processor, which avoids 
expensive data copies of traditional block-based processors and avoids materialization of adjacency lists in 
blocks, a new data structure we call single-indexed property pages and an accompanying edge ID scheme, and a new application of Jacobson's bit vector index for compressing NULL and empty lists.  

\iflongversion
\begin{acks}
This work was supported in part by an NSERC Discovery grant. We thank Lori Paniak for promptly assisting with many system issues. We also thank Xiyang Feng, Guodong Jin and Siddhartha Sahu for helping at different stages of this project and Snehal Mishra for helping with illustrations. We thank the
anonymous reviewers for their valuable comments.
\end{acks}
\fi


\bibliographystyle{ACM-Reference-Format}
\bibliography{references}
\iflongversion
\begin{appendix}

\section{Sensitivity Analyses}

\label{app:sensitivity}
Two of our data structures have parameters that can be modified. First is 
the parameter $k$, which groups the properties of edges in $k$ many adjacency lists into a single property page.
The other are the $m$ and $c$ parameters in our Jacobson's index-based NULL compression schemes.
In this section, we do a sensitivity analysis on these parameters to demonstrate their effects.

\subsection{Parameter $k$}
\label{subsec:sensitivity-k}

In this experiment, we extend our experiment from Section~\ref{exp:property-pages},
where we ran 1- and 2-hop queries with a predicate on the edges 
on \texttt{LDBC100}, \texttt{WIKI}, and \texttt{FLICKR},
comparing property pages with $k=128$ parameter and pure edge columns. Note
the pure edge columns store all of the edges in a single column, so is equivalent to setting
$k=\infty$. So we repeat the same experiment in Table~\ref{tbl:prop-pages} with $k=2^i$ for $i=1, ..., 17$.
As before, we measure the runtime performance of the query, as a proxy for the efficiency of accessing 
the properties, which is the main computation done in these queries. Our results are shown 
in Figure~\ref{fig:senstivity-k}. The last x-axis value ``*'' uses pure edge columns (so corresponds 
to the COL$_E$ values in Table~\ref{tbl:prop-pages}). Note that for most settings (except LDBC 2H figure)
using up to $k=2^9$ yields relatively stable results, after which the performance 
degrades.
This threshold value is a bit larger, up to $k=2^{11}$ on 
\texttt{Flickr}, which has a lower average degree than \texttt{LDBC} and \texttt{Wiki} (14 vs 44 and 41, respectively).
This is expected because in this experiment, we read all of the edge properties in the order of the adjacency lists.
We expect there to be some threshold property page block size $B$, under which we expect to get 
good cache locality. If the sizes of the property pages get larger than $B$ we expect to lose locality and
the performance to degrade. Therefore the smaller the average adjacency list sizes, the larger number of 
adjacency lists we can pack together into $B$ to still get good cache locality. In light of this analysis, our choice
of $k=2^7=128$ is in the safe region in all of these settings. Recall also that, although we do not 
focus on updates in this paper, making $k$ very small, say 2 or 4, is not a good choice, as it would make recycling 
deleted positional offsets, which necessarily leave gaps, difficult. 

\begin{figure}[h!]
	\centering
	\includegraphics[width=0.9\columnwidth]{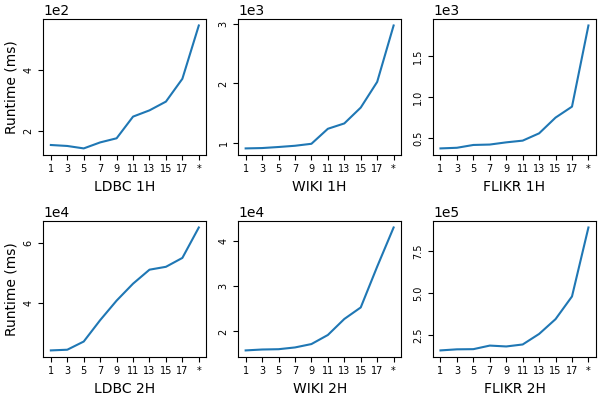}
	\captionsetup{justification=centering}
	\caption{Sensitivity analysis of k. Values of X-axis are exponents of 2 which denotes parameter k of the property pages. '*' denotes Edge Column which is essentially a property page with k = $\infty$}
	\label{fig:senstivity-k}
	\vspace{-10pt}
\end{figure}

\subsection{Parameters $m$ and $c$}
\label{subsec:sensitivity-m-c}
We next analyze the effects of $m$ and $c$ in our NULL compression scheme
that extends Abadi's bitstring-based scheme with the Jacobson's index. Recall
that the parameter $c$ determines the size of the pre-populated map to compute ranks. 
However in practice this can be only 8 or 16, because while at $c=8$ and 
$c=16$ the map sizes are modest at 2KB and 1MB, at $c=24$ and $c=32$ the size increases exponentially and 
becomes very large: 
1.6GB and 546GB. Therefore, we only consider map with size either $c=8$ or $c=16$.
Recall that the parameter $m$ denotes the number of bytes we use to store the prefix sums. This parameter
can take on a larger range of values, such as 8, 16, 24, 32 or even larger.
We expect m to primarily affect the memory overhead of our scheme and not the decompression time.
Specifically our memory
overhead per element compressed is exactly $m/c$ bits. Therefore, for any $c$ value, there is value in picking
a small $m$. 

To see if there is any visible decompression time differences between $c=8$ and $c=16$
and to verify that different $m$ values do not have any effect on performance, we 
reran the experiment from Section~\ref{exp:prefixSum} with different (c,m) combinations,
where we varied $m$ to be 8, 16, 24, or 32. 
As before we measure the runtime of the 
MATCH (a:\texttt{Person})$-$[e:\texttt{Likes}]$\rightarrow$ (b:\texttt{Comment}) RETURN b.\texttt{creationDate}
query where \texttt{creationDate} property contains different percentage of NULL values.
Our runtime results are shown in Table~\ref{tbl:senstivity-m-c-runtime}. In the table, $\rho$
is the percentage of non-NULL values in the \texttt{creationDate} property. As we expect, we do not 
see any visible sensitivity to $m$, at any non-NULL percentage value $\rho$. Similarly we also do not see any visible sensitivity between $c=8$ vs $c=16$.

Our memory numbers are in Table~\ref{tbl:senstivity-m-c-storage}. We report only our overheads due to
storing the prefix sums and the bitmaps and not the non-NULL values. Since these overheads
are fixed for any $\rho$ value, this table has a single row (measured at $\rho$=50\%). 
As we expect the measured overheads are determined by the $m/c$ value. For example,
(8, 8) and (16, 16) have almost identical measured overheads,  52.7MB and 52.8MB, respectively.  
We think (8, 8), (16, 16), or (16, 8) are reasonable choices in practice for (c, m) as they achieve the lowest
overheads.
 
\begin{table}[h!]
	\centering
	\bgroup
	\setlength{\tabcolsep}{5pt}
	\def\arraystretch{1}
	\begin{tabular}{ |c|c|c|c|c|c|c|c|c| } 
		\hline
		$\rho$ & 8,8 & 8,16 & 8,24 & 8,32 & 16,8 & 16,16 & 16,24 & 16,32 \\
		\hline
		\hline
		\texttt{100} & 21.0 & 20.9 & 20.2  & 20.3 & 20.1 & 20.6 & 20.1 & 20.8 \\
		\texttt{90}  & 20.9 & 20.2 & 20.5  & 20.0 & 20.0 & 20.5 & 20.7 & 20.9 \\
		\texttt{80}  & 20.3 & 20.1 & 20.2  & 20.1 & 20.1 & 21.0 & 20.9 & 20.4 \\
		\texttt{70}  & 19.0 & 19.8 & 19.8  & 20.0 & 19.8 & 19.7 & 20.4 & 20.3 \\ 
		\texttt{60}  & 19.0 & 19.0 & 20.2  & 19.2 & 19.0 & 19.1 & 19.6 & 19.2 \\
		\texttt{50}  & 18.2 & 18.7 & 18.6  & 18.4 & 18.8 & 18.1 & 18.5 & 18.8 \\ 
		\texttt{40}  & 17.2 & 16.5 & 16.3  & 16.1 & 16.7 & 17.7 & 17.9 & 16.9 \\
		\texttt{30}  & 16.9 & 14.4 & 14.6  & 14.4 & 14.8 & 15.9 & 15.6 & 15.6 \\
		\texttt{20}  & 13.7 & 12.2 & 12.1  & 12.2 & 12.2 & 13.6 & 13.9 & 13.0 \\
		\texttt{10}  & 10.9 & 10.2 & 10.2  & 10.3 & 10.3 & 11.0 & 10.4 & 10.8 \\
		\cline{2-7} 
		\hline
	\end{tabular}
	\egroup
	\captionsetup{justification=centering}
	\caption{Sensitivity analysis of (c,m): runtime in ms. $\rho$ is the percetage of non-NULL values in the \texttt{creationDate} property.}
	\label{tbl:senstivity-m-c-runtime}
	\vspace{-5pt}
\end{table}

\begin{table}[h!]
	\centering
	\bgroup
	\setlength{\tabcolsep}{5pt}
	\def\arraystretch{1}
	\begin{tabular}{ |c|c|c|c|c|c|c|c| } 
		\hline
		8,8 & 8,16 & 8,24 & 8,32 & 16,8 & 16,16 & 16,24 & 16,32 \\
		\hline
		\hline
		52.8 & 78.8 & 105.2 & 132.6 & 39.6 & 52.7 & 65.2 & 80.1 \\
		\cline{2-6} 
		\hline
	\end{tabular}
	\egroup
	\captionsetup{justification=centering}
	\caption{Sensitivity analysis of (c,m): overhead (in MB) of storing null bits and prefixSums for different configurations of c and m values.}
	\label{tbl:senstivity-m-c-storage}
	\vspace{-5pt}
\end{table}

\section{Modified LDBC SNB Queries}
\label{app:ldbc-snb}


\begin{ldbcQ}
IS01
{\em 
\begin{lstlisting}[numbers=none,  showstringspaces=false,belowskip=0pt ]
MATCH  (p:Person)$-$[:IsLocatedIn]$\rightarrow$(pl:Place)
WHERE  p.id = 22468883 
RETURN p.fName, p.lName, p.birthday, p.locationIP, 
       p.browserUsed, p.gender, p.creationDate,
       pl.id;
\end{lstlisting}
}
\end{ldbcQ}

\begin{ldbcQ}
IS02
{\em 
\begin{lstlisting}[numbers=none,
showstringspaces=false,belowskip=0pt ]
MATCH  (p:Person)$\leftarrow$[:hasCreator]$-$(c:Comment)
       (c:Comment)$-$[:replyOf]$\rightarrow$(post:Post)
       (post:Post)$-$[:hasCreator]$\rightarrow$(op:Person)
WHERE  p.id = 22468883 
RETURN c.id, c.content, c.creationDate, op.id, 
       op.fName, op.lName;
\end{lstlisting}
}
\end{ldbcQ}

\begin{ldbcQ}
IS03
{\em 
\begin{lstlisting}[numbers=none,  showstringspaces=false,belowskip=0pt ]
MATCH  (p:Person)$-$[k:knows]$\rightarrow$(friend:Person)
WHERE  p.id = 22468883 
RETURN friend.id, friend.fName, friend.lName, 
       e.date;
\end{lstlisting}
}
\end{ldbcQ}

\begin{ldbcQ}
IS04
{\em 
\begin{lstlisting}[numbers=none,  showstringspaces=false,belowskip=0pt ]
MATCH  (comment:Comment)
WHERE  comment.id = 0 
RETURN comment.creationDate, comment.content;
\end{lstlisting}
}
\end{ldbcQ}

\begin{ldbcQ}
IS05
{\em 
\begin{lstlisting}[numbers=none,  showstringspaces=false,belowskip=0pt ]
MATCH  (comment:Comment)$-$[:hasCreator]$\rightarrow$(p:Person)
WHERE  comment.id = 0 
RETURN p.id, p.fName, p.lName;
\end{lstlisting}
	}
\end{ldbcQ}

\begin{ldbcQ}
IS06
{\em 
\begin{lstlisting}[numbers=none,  showstringspaces=false,belowskip=0pt ]
MATCH  (comment:Comment)$-$[:replyOf]$\rightarrow$(pst:Post)
       (pst:Post)$\leftarrow$[:containerOf]$-$(f:Forum)
       (f:Forum)$-$[:hasModerator]$\rightarrow$(p:Person)
WHERE  comment.id = 0
RETURN f.id, f.title, p.id, p.fName, p.lName;
\end{lstlisting}
}
\end{ldbcQ}

\begin{ldbcQ}
IS07
{\em 
\begin{lstlisting}[numbers=none,  showstringspaces=false,belowskip=0pt ]
MATCH  (mAuth:Person)$\leftarrow$[:hasCreator]$-$(cmt0:Comment)
       (cmt0:Comment)$\leftarrow$[:replyOf]$-$(cmt1:Comment)
       (cmt1:Comment)$-$[:hasCreator]$\rightarrow$(rAuth:Person)
WHERE  comment.id = 6
RETURN cmt1.id, cmt1.content, cmt1.creationDate, 
       rAuth.id, rAuth.fName, rAuth.lName;
\end{lstlisting}
}
\end{ldbcQ}

\begin{ldbcQ}
IC01
{\em 
\begin{lstlisting}[numbers=none,  showstringspaces=false,belowskip=0pt ]
MATCH  (person:Person)$-$[:knows]$\rightarrow$(p1:Person)
       (p1:Person)$-$[:knows]$\rightarrow$(p2:Person)
       (p2:Person)$-$[:knows]$\rightarrow$(op:Person)
       (op:Person)$-$[:isLocatedIn]$\rightarrow$(pl:Place)
WHERE  person.id = 22468883
RETURN op.id, op.lName, op.birthday, op.creationDate, 
       op.gender, op.locationIP, city.name;
\end{lstlisting}
}
\end{ldbcQ}

\begin{ldbcQ}
IC02
{\em 
\begin{lstlisting}[numbers=none,  showstringspaces=false,belowskip=0pt ]
MATCH  (p:Person)$-$[:knows]$\rightarrow$(frnd:Person)
       (frnd:Person)$\leftarrow$[:hasCreator]$-$(msg:Comment)
WHERE  p.id = 22468883 AND 
       msg.creationDate < 1342805711
RETURN frnd.id, frnd.fName, frnd.lName, msg.id, 
       msg.content,msg.creationDate;
\end{lstlisting}
}
\end{ldbcQ}

\begin{ldbcQ}
IC03
{\em 
\begin{lstlisting}[numbers=none,  showstringspaces=false,belowskip=0pt ]
MATCH  (person:Person)$-$[:knows]$\rightarrow$(p1:Person)
       (p1:Person)$-$[:knows]$\rightarrow$(op:Person)
       (op:Person)$-$[:isLocatedIn]$\rightarrow$(pl:Place)
       (op:Person)$\leftarrow$[:hasCreator]$-$(mx:Comment)
       (mx:Comment)$-$[:isLocatedIn]$\rightarrow$(px:Place)
       (op:Person)$\leftarrow$[:hasCreator]$-$(my:Comment)
       (my:Comment)$-$[:isLocatedIn]$\rightarrow$(py:Place)
WHERE  person.id = 22468883 AND
       mx.creationDate >= 1313591219 AND 
       mx.creationDate <= 1513591219 AND
       my.creationDate >= 1313591219 AND 
       my.creationDate <= 1513591219 AND
       cx.name = 'India' AND cy.name = 'China'
RETURN cmt1.id, cmt1.content, cmt1.creationDate, 
       rAuth.id, rAuth.fName, rAuth.lName;
\end{lstlisting}
}
\end{ldbcQ}

\begin{ldbcQ}
IC04
{\em 
\begin{lstlisting}[numbers=none,  showstringspaces=false,belowskip=0pt ]
MATCH  (:Person)$\leftarrow$[:knows]$-$(p:Person)
       (p:Person)$-$[:knows]$\rightarrow$(frnd:Person)
       (frnd:Person)$\leftarrow$[:hasCreator]$-$(pst:Post)
       (pst:Post)$-$[:hasTag]$\rightarrow$(t:Tag)
WHERE  p.id = 22468883 AND
       post.creationDate >= 1313591219 AND 
       post.creationDate <= 1513591219
RETURN cmt1.id, cmt1.content, cmt1.creationDate, 
       rAuth.id, rAuth.fName, rAuth.lName;
\end{lstlisting}
}
\end{ldbcQ}

\begin{ldbcQ}
IC05
{\em 
\begin{lstlisting}[numbers=none,  showstringspaces=false,belowskip=0pt ]
MATCH  (p1:Person)$-$[:knows]$\rightarrow$(p2:Person)
       (p2:Person)$-$[:knows]$\rightarrow$(p3:Person)
       (p3:Person)$\leftarrow$[hm:hasMember]$-$(f:Forum)
       (f:Forum)$-$[:containerOf]$\rightarrow$(pst:Post)
WHERE  p1.id = 22468883 AND hm.date > 1267302820
RETURN f.title;
\end{lstlisting}
}
\end{ldbcQ}

\begin{ldbcQ}
IC06
{\em 
\begin{lstlisting}[numbers=none,  showstringspaces=false,belowskip=0pt ]
MATCH  (p1:Person)$-$[:knows]$\rightarrow$(p2:Person)
       (p2:Person)$-$[:knows]$\rightarrow$(p3:Person)
       (p3:Person)$\leftarrow$[:hasCreator]$-$(pst:Post)
       (pst:Post)$-$[:hasTag]$\rightarrow$(t1:Tag)
       (pst:Post)$-$[:hasTag]$\rightarrow$(t2:Tag)
WHERE  p1.id = 22468883 AND 
       t1.name = 'Rumi' AND t2.name <> 'Rumi'
RETURN t2.name;
\end{lstlisting}
}
\end{ldbcQ}

\begin{ldbcQ}
IC07
{\em 
\begin{lstlisting}[numbers=none,  showstringspaces=false,belowskip=0pt ]
MATCH  (p:Person)$\leftarrow$[:hasCreator]$-$(cmt:Comment)
       (cmt:Comment)$\leftarrow$[l:likes]$-$(frnd:Person)
WHERE  p.id = 22468883
RETURN frnd.id, frnd.fName, frnd.lName, l.date, 
       cmt.content;
\end{lstlisting}
}
\end{ldbcQ}

\begin{ldbcQ}
IC08
{\em 
\begin{lstlisting}[numbers=none,  showstringspaces=false,belowskip=0pt ]
MATCH  (p:Person)$\leftarrow$[:hasCreator]$-$(pst:Post)
       (pst:Post)$\leftarrow$[:replyOf]$-$(cmt:Comment)
       (cmt:Comment)$-$[:hasCreator]$\rightarrow$(cmtAuth:Person)
WHERE  p1.id = 22468883
RETURN cmtAuth.id, cmtAuth.fName, cmtAuth.lName, 
       cmt.creationDate, cmt.id, cmt.content;
\end{lstlisting}
}
\end{ldbcQ}

\begin{ldbcQ}
IC09
{\em 
\begin{lstlisting}[numbers=none,  showstringspaces=false,belowskip=0pt ]
MATCH  (p1:Person)$-$[:knows]$\rightarrow$(p2:Person)
       (p2:Person)$-$[:knows]$\rightarrow$(p3:Person)
       (p3:Person)$\leftarrow$[:hasCreator]$-$(cmt:Comment)
WHERE  person.id = 22468883 AND 
       cmt.creationDate < 1342840042
RETURN p3.id, p3.fName, p3.lName, cmt.id, cmt.content,
       cmt.creationDate;
\end{lstlisting}
}
\end{ldbcQ}

\begin{ldbcQ}
IC11
{\em 
\begin{lstlisting}[numbers=none,  showstringspaces=false,belowskip=0pt ]
MATCH  (p1:Person)$-$[:knows]$\rightarrow$(p2:Person)
       (p2:Person)$-$[:knows]$\rightarrow$(p3:Person)
       (p3:Person)$-$[w:workAt]$\rightarrow$(org:Organization)
       (org:Organization)$-$[:isLocatedIn]$\rightarrow$(pl:Place)
WHERE  p1.id = 22468883 AND w.year < 2016 AND 
       pl.name = 'China'
RETURN p3.id, p3.fName, p3.lName, org.name;
\end{lstlisting}
}
\end{ldbcQ}

\begin{ldbcQ}
IC12
{\em 
\begin{lstlisting}[numbers=none,  showstringspaces=false,belowskip=0pt ]
MATCH  (p1:Person)$-$[:knows]$\rightarrow$(p2:Person)
       (p2:Person)$\leftarrow$[:hasCreator]$-$(cmt:Comment)
       (cmt:Comment)$-$[:replyOf]$-$(pst:Post)
       (pst:Post)$-$[:hasTag]$\rightarrow$(t:Tag)
       (t:Tag)$-$[:hasType]$\rightarrow$(tc:TagClass)
       (tc:TagClass)$-$[:isSubclassOf]$\rightarrow$(:TagClass)
WHERE  p1.id = 22468883 AND tc.name='Person'
RETURN p2.id, p2.fName, p2.lName;
\end{lstlisting}
}
\end{ldbcQ}

\section{Modified JOB Queries}
\label{app:job-benchmark-queries}
	
\begin{jobQ}
1A
{\em 
\begin{lstlisting}[numbers=none,showstringspaces=false,belowskip=0pt ]
MATCH  (t:title)$-$[mc:movie_companies]$\rightarrow$(:company_name), 
       (t:title)$-$[:has_mov_info_2]$\rightarrow$(mii:mov_info_2)
WHERE  mc.company_type = 'production company' AND 
       mc.note CONTAINS '(co-production)' AND
       mii.info_type = 'top 250 rank'
RETURN COUNT(*);
\end{lstlisting}
}
\end{jobQ}
	
\begin{jobQ}
2A
{\em 
\begin{lstlisting}[numbers=none,showstringspaces=false,belowskip=0pt ]
MATCH  (t:title)$-$[:movie_companies]$\rightarrow$(cn:company_name), 
       (t:title)$-$[:movie_keyword]$\rightarrow$(k:keyword)
WHERE  cn.country_code = '[de]' AND 
       k.keyword = 'character-name-in-title'
RETURN COUNT(*);
\end{lstlisting}
}
\end{jobQ}

\begin{jobQ}
3A
{\em 
\begin{lstlisting}[numbers=none,  showstringspaces=false,belowskip=0pt ]
MATCH  (t:title)$-$[:movie_keyword]$\rightarrow$(k:keyword), 
       (t:title)$-$[:has_movie_info]$\rightarrow$(mi:movie_info)
WHERE  t.production_year > 2005 AND
       k.keyword CONTAINS 'sequel' AND 
       mi.info = 'Sweden'
RETURN COUNT(*);

\end{lstlisting}
}
\end{jobQ}
	
\begin{jobQ}
4A
{\em 
\begin{lstlisting}[numbers=none,  showstringspaces=false,belowskip=0pt ]
MATCH  (t:title)$-$[:movie_keyword]$\rightarrow$(k:keyword),  
       (t:title)$-$[:has_mov_info_2]$\rightarrow$(mii:mov_info_2)
WHERE  t.production_year > 2005 AND
       k.keyword CONTAINS 'sequel' AND 
       mii.info_type = 'rating' AND mii.info > '5.0' 
RETURN COUNT(*);
\end{lstlisting}
}
\end{jobQ}
	
\begin{jobQ}
5A
{\em 
\begin{lstlisting}[numbers=none,  showstringspaces=false,belowskip=0pt ]
MATCH  (t:title)$-$[mc:movie_companies]$\rightarrow$(:company_name), 
       (t:title)$-$[:has_movie_info]$\rightarrow$(:movie_info)
WHERE  t.production_year > 2005 AND
       mc.company_type = 'production company' AND 
       mc.note CONTAINS '(theatrical)' AND 
       mc.note CONTAINS '(France)'
RETURN COUNT(*);
\end{lstlisting}
}
\end{jobQ}
	
\begin{jobQ}
6A
{\em 
\begin{lstlisting}[numbers=none,  showstringspaces=false,belowskip=0pt ]
MATCH  (t:title)$-$[:cast_info]$\rightarrow$(n:name), 
       (t:title)$-$[:movie_keyword]$\rightarrow$(k:keyword)
WHERE  t.production_year > 2010 AND
       n.name CONTAINS 'Downey' AND
       k.keyword = 'marvel-cinematic-universe'
RETURN COUNT(*);
\end{lstlisting}
}
\end{jobQ}
	
\begin{jobQ}
7A
{\em 
\begin{lstlisting}[numbers=none,  showstringspaces=false,belowskip=0pt ]
MATCH  (t:title)$-$[ml:movie_link]$\rightarrow$(:title), 
       (t:title)$-$[:cast_info]$\rightarrow$(n:name), 
       (n:name)$-$[:has_aka_name]$\rightarrow$(an:aka_name),
       (n:name)$-$[:has_person_info]$\rightarrow$(pi:person_info)
WHERE  t.production_year >= 1980 AND
       t.production_year <= 1995 AND 
       ml.link_type = 'features' AND
       n.name_pcode_cf >= 'A' AND
       n.name_pcode_cf <= 'F' AND n.gender = 'm' AND
       an.name CONTAINS 'a' AND 
       pi.info_type = 'mini biography' AND  
       pi.note = 'Volker Boehm'
RETURN COUNT(*);
\end{lstlisting}
}
\end{jobQ}
	
\begin{jobQ}
8A
{\em 
\begin{lstlisting}[numbers=none,  showstringspaces=false,belowskip=0pt ]
MATCH  (t:title)$-$[mc:movie_companies]$\rightarrow$(cn:company_name), 
       (t:title)$-$[ci:cast_info]$\rightarrow$(n:name), 
       (n:name)$-$[:has_aka_name]$\rightarrow$(:aka_name)
WHERE  mc.note CONTAINS '(Japan)' AND
       cn.country_code ='[jp]' AND
       ci.note ='(voice: English version)' AND 
       ci.role = 'actress' AND n.name CONTAINS 'Yo'
RETURN COUNT(*);
\end{lstlisting}
}
\end{jobQ}
	
\begin{jobQ}
9A
{\em 
\begin{lstlisting}[numbers=none,  showstringspaces=false,belowskip=0pt]
MATCH  (t:title)$-$[mc:movie_companies]$\rightarrow$(cn:company_name), 
       (t:title)$-$[ci:cast_info]$\rightarrow$(n:name), 
       (n:name)$-$[:has_aka_name]$\rightarrow$(:aka_name)
WHERE  t.production_year >= 2005 AND 
       t.production_year <= 2015 AND
       mc.note CONTAINS '(USA)' AND 
       cn.country_code = '[us]' AND 
       ci.role = 'actress' AND
       ci.note STARTS WITH '(voice' AND
       n.gender = 'f' AND n.name CONTAINS 'Ang'
RETURN COUNT(*);
\end{lstlisting}
}
\end{jobQ}
	
\begin{jobQ}
10A
{\em 
\begin{lstlisting}[numbers=none,  showstringspaces=false,belowskip=0pt ]
MATCH  (t:title)$-$[:movie_companies]$\rightarrow$(cn:company_name), 
       (t:title)$-$[ci:cast_info]$\rightarrow$(:name)
WHERE  t.production_year > 2005 AND
       cn.country_code = '[ru]' AND
       ci.note CONTAINS '(uncredited)' AND
       ci.note CONTAINS '(voice)' AND ci.role = 'actor'
RETURN COUNT(*);
\end{lstlisting}
}
\end{jobQ}
	
\begin{jobQ}
11A
{\em 
\begin{lstlisting}[numbers=none,  showstringspaces=false,belowskip=0pt ]
MATCH  (t:title)$-$[ml:movie_link]$\rightarrow$(:title), 
       (t:title)$-$[mc:movie_companies]$\rightarrow$(cn:company_name),
       (t:title)$-$[:movie_keyword]$\rightarrow$(k:keyword)
WHERE  t.production_year > 1950 AND 
       t.production_year < 2000 AND 
       ml.link_type IN ('follows', 'followedBy') AND 
       ml.link_type LIKE '%follow%' AND
       mc.company_type = 'production company' AND 
       cn.country_code <> '[pl]' AND 
       cn.name CONTAINS 'Film' AND k.keyword ='sequel'
RETURN COUNT(*);
\end{lstlisting}
}
\end{jobQ}
	
\begin{jobQ}
12A
{\em 
\begin{lstlisting}[numbers=none,  showstringspaces=false,belowskip=0pt ]
MATCH  (t:title)$-$[:has_movie_info]$\rightarrow$(mi:movie_info), 
       (t:title)$-$[mc:movie_companies]$\rightarrow$(cn:company_name),
       (t:title)$-$[:has_mov_info_2]$\rightarrow$(mii:mov_info_2)
WHERE  t.production_year >= 2005 AND 
       t.production_year <= 2008 AND
       mii.info > '8.0' AND 
       mi.info_type = 'genres' AND
       mi.info = 'Drama' AND 
       mc.company_type = 'production company' AND 
       cn.country_code = '[us]' AND 
       mii.info_type = 'rating'
RETURN COUNT(*);
\end{lstlisting}
}
\end{jobQ}
	
\begin{jobQ}
13A
{\em 
\begin{lstlisting}[numbers=none,  showstringspaces=false,belowskip=0pt ]
MATCH  (t:title)$-$[:has_movie_info]$\rightarrow$(mi:movie_info), 
       (t:title)$-$[mc:movie_companies]$\rightarrow$(cn:company_name),
       (t:title)$-$[:has_mov_info_2]$\rightarrow$(mii:mov_info_2)
WHERE  t.kind = 'movie' AND 
       mi.info_type = 'release dates' AND
       mc.company_type = 'production company' AND
       cn.country_code = '[de]' AND 
       mii.info_type = 'rating'
RETURN COUNT(*);
\end{lstlisting}
}
\end{jobQ}
	
\begin{jobQ}
14A
{\em 
\begin{lstlisting}[numbers=none,  showstringspaces=false,belowskip=0pt ]
MATCH  (t:title)$-$[:has_movie_info]$\rightarrow$(mi:movie_info), 
       (t:title)$-$[:movie_keyword]$\rightarrow$(k:keyword),
       (t:title)$-$[:has_mov_info_2]$\rightarrow$(mii:mov_info_2)
WHERE  t.production_year > 2010 AND 
       t.kind = 'movie' AND mi.info = 'USA' AND 
       mi.info_type = 'countries' AND 
       k.keyword = 'murder' AND  mii.info < '8.5' AND 
       mii.info_type = 'rating'
RETURN COUNT(*);
\end{lstlisting}
}
\end{jobQ}
	
\begin{jobQ}
15A
{\em 
\begin{lstlisting}[numbers=none,  showstringspaces=false,belowskip=0pt ]
MATCH  (t:title)$-$[:has_movie_info]$\rightarrow$(mi:movie_info), 
       (t:title)$-$[mc:movie_companies]$\rightarrow$(cn:company_name),
       (t:title)$-$[:movie_keyword]$\rightarrow$(:keyword)
WHERE  t.production_year > 2000 AND 
       mi.info STARTS WITH 'USA:' AND 
       mi.note CONTAINS 'internet' AND 
       mi.info_type = 'release dates' AND
       mc.note CONTAINS '(worldwide)' AND  
       mc.note CONTAINS '(200' AND 
       cn.country_code = '[us]'
RETURN COUNT(*);
\end{lstlisting}
}
\end{jobQ}
	
\begin{jobQ}
16A
{\em 
\begin{lstlisting}[numbers=none,  showstringspaces=false,belowskip=0pt ]
MATCH  (t:title)$-$[:movie_keyword]$\rightarrow$(k:keyword), 
       (t:title)$-$[:movie_companies]$\rightarrow$(cn:company_name), 
       (t:title)$-$[:cast_info]$\rightarrow$(:name),
       (n:name)$-$[:has_aka_name]$\rightarrow$(:aka_name)
WHERE  t.episode_nr >= 50 AND t.episode_nr < 100 AND 
       k.keyword = 'character-name-in-title' AND   
       cn.country_code = '[us]'
RETURN COUNT(*);
\end{lstlisting}
}
\end{jobQ}
	
\begin{jobQ}
17A
{\em 
\begin{lstlisting}[numbers=none,  showstringspaces=false,belowskip=0pt ]
MATCH  (t:title)$-$[:cast_info]$\rightarrow$(n:name), 
       (t:title)$-$[:movie_companies]$\rightarrow$(cn:company_name), 
       (t:title)$-$[:movie_keyword]$\rightarrow$(k:keyword)
WHERE  n.name STARTS WITH 'B' AND
       cn.country_code ='[us]' AND 
       k.keyword ='character-name-in-title'
RETURN COUNT(*);
\end{lstlisting}
}
\end{jobQ}
	
\begin{jobQ}
18A
{\em 
\begin{lstlisting}[numbers=none,  showstringspaces=false,belowskip=0pt ]
MATCH  (t:title)$-$[:has_movie_info]$\rightarrow$(mi:movie_info), 
       (t:title)$-$[:has_mov_info_2]$\rightarrow$(mii:mov_info_2),
       (t:title)$-$[:cast_info]$\rightarrow$(n:name)
WHERE  mi.info_type = 'budget' AND 
       mii.info_type = 'votes' AND
       n.name CONTAINS 'Tim' AND n.gender = 'm' 
RETURN COUNT(*);
\end{lstlisting}
		}
	\end{jobQ}

\begin{jobQ}
19A
{\em 
\begin{lstlisting}[numbers=none,  showstringspaces=false,belowskip=0pt ]
MATCH  (t:title)$-$[:has_movie_info]$\rightarrow$(mi:movie_info), 
       (t:title)$-$[mc:movie_companies]$\rightarrow$(cn:company_name),
       (t:title)$-$[ci:cast_info]$\rightarrow$(n:name),
       (n:name)$-$[:has_aka_name]$\rightarrow$(:aka_name)
WHERE  t.production_year >= 2005 AND
       t.production_year <= 2009 AND
       mi.info_type = 'release dates' AND 
       mi.info STARTS WITH 'Japan:' AND
       mc.note CONTAINS '(USA)' AND 
       cn.country_code = '[us]' AND 
       ci.note STARTS WITH '(voice' AND 
       n.gender = 'f' AND ci.role = 'actress' AND 
       n.name CONTAINS 'Ang'
RETURN COUNT(*);
\end{lstlisting}
}
\end{jobQ}

\begin{jobQ}
20A
{\em 
\begin{lstlisting}[numbers=none,  showstringspaces=false,belowskip=0pt ]
MATCH  (t:title)$-$[:movie_keyword]$\rightarrow$(k:keyword), 
       (t:title)$-$[:has_complete_cast]$\rightarrow$(cc:complete_cast), 
       (t:title)$-$[ci:cast_info]$\rightarrow$(:name)
WHERE  t.production_year > 1950 AND
       t.kind = 'movie' AND 
       k.keyword = 'superhero' AND 
       cc.subject = 'cast' AND 
       cc.status IN ('complete', 'complete+verified') AND 
       ci.name CONTAINS 'Tony' AND 
       ci.name CONTAINS 'Stark'
RETURN COUNT(*);
\end{lstlisting}
}
\end{jobQ}

\begin{jobQ}
21A
{\em 
\begin{lstlisting}[numbers=none,  showstringspaces=false,belowskip=0pt ]
MATCH  (t:title)$-$[:has_movie_info]$\rightarrow$(mi:movie_info), 
       (t:title)$-$[mc:movie_companies]$\rightarrow$(cn:company_name),
       (t:title)$-$[:movie_keyword]$\rightarrow$(k:keyword),
       (t:title)$-$[ml:movie_link]$\rightarrow$(:title)
WHERE  t.production_year >= 1950 AND 
       t.production_year <= 2000 
       mi.info = 'Germany' AND
       mc.company_type = 'production company' AND
       cn.country_code <> '[pl]' AND 
       cn.name CONTAINS 'Film' AND 
       k.keyword CONTAINS 'sequel' AND
       ml.link_type IN ('follows', 'followedBy')
RETURN COUNT(*);
\end{lstlisting}
}
\end{jobQ}

\begin{jobQ}
22A
{\em 
\begin{lstlisting}[numbers=none,  showstringspaces=false,belowskip=0pt ]
MATCH  (t:title)$-$[:has_movie_info]$\rightarrow$(mi:movie_info), 
       (t:title)$-$[:has_mov_info_2]$\rightarrow$(mii:mov_info_2),
       (t:title)$-$[mc:movie_companies]$\rightarrow$(cn:company_name),
       (t:title)$-$[:movie_keyword]$\rightarrow$(k:keyword)
WHERE  t.production_year > 2008 AND 
       t.kind = 'movie' AND mi.info = 'USA' AND 
       mi.info_type = 'countries' AND
       mii.info_type = 'rating' AND 
       mii.info < '7.0' AND 
       mc.note CONTAINS '(200' AND 
       cn.country_code <> '[us]' AND 
       k.keyword = 'murder' 
RETURN COUNT(*);
\end{lstlisting}
}
\end{jobQ}

\begin{jobQ}
23A
{\em 
\begin{lstlisting}[numbers=none,  showstringspaces=false,belowskip=0pt ]
MATCH  (t:title)$-$[:has_movie_info]$\rightarrow$(mi:movie_info), 
       (t:title)$-$[:movie_companies]$\rightarrow$(cn:company_name),
       (t:title)$-$[:movie_keyword]$\rightarrow$(:keyword), 
       (t:title)$-$[:has_complete_cast]$\rightarrow$(cc:complete_cast)
WHERE  t.production_year > 2000 AND 
       t.kind = 'movie' AND  
       mi.info_type = 'release dates' AND 
       mi.note CONTAINS 'internet' AND
       mi.info STARTS WITH 'USA:' AND 
       cn.country_code = '[us]' AND 
       cc.status = 'complete+verified'
RETURN COUNT(*);
\end{lstlisting}
	}
\end{jobQ}

\begin{jobQ}
24A
{\em 
\begin{lstlisting}[numbers=none,  showstringspaces=false,belowskip=0pt ]
MATCH  (t:title)$-$[:has_movie_info]$\rightarrow$(mi:movie_info), 
       (t:title)$-$[:movie_companies]$\rightarrow$(cn:company_name),
       (t:title)$-$[ci:cast_info]$\rightarrow$(n:name), 
       (n:name)$-$[:has_aka_name]$\rightarrow$(:aka_name),
       (t:title)$-$[:movie_keyword]$\rightarrow$(k:keyword)
WHERE  t.production_year > 2010 AND 
       mi.info_type = 'release dates' AND
       mi.info STARTS WITH 'USA:' AND
       cn.country_code = '[us]' AND 
       ci.note STARTS WITH '(voice:' AND 
       ci.role = 'actress' AND n.gender = 'f'  AND  
       k.keyword = 'hero'        
RETURN COUNT(*);
\end{lstlisting}
	}
\end{jobQ}

\begin{jobQ}
25A
{\em 
\begin{lstlisting}[numbers=none,  showstringspaces=false,belowskip=0pt ]
MATCH  (t:title)$-$[:has_movie_info]$\rightarrow$(mi:movie_info), 
       (t:title)$-$[:has_mov_info_2]$\rightarrow$(mii:mov_info_2),
       (t:title)$-$[:movie_keyword]$\rightarrow$(k:keyword), 
       (t:title)$-$[:cast_info]$\rightarrow$(n:name)
WHERE  mi.info_type = 'genres' AND 
       mii.info_type = 'votes' AND        
       k.keyword = 'murder' AND mi.info = 'Horror' AND
       n.gender = 'm' 
RETURN COUNT(*);
\end{lstlisting}
	}
\end{jobQ}

\begin{jobQ}
26A
{\em 
\begin{lstlisting}[numbers=none,  showstringspaces=false,belowskip=0pt ]
MATCH  (t:title)$-$[:has_mov_info_2]$\rightarrow$(mii:mov_info_2),  
       (t:title)$-$[:movie_keyword]$\rightarrow$(k:keyword),
       (t:title)$-$[ci:cast_info]$\rightarrow$(:name), 
       (t:title)$-$[:has_complete_cast]$\rightarrow$(cc:complete_cast)
WHERE  t.production_year > 2000 AND 
       t.kind = 'movie' AND mii.info > '7.0' AND
       mii.info_type = 'rating' AND 
       k.keyword = 'superhero' AND 
       ci.name CONTAINS 'man' AND 
       cc.subject = 'cast' AND 
       cc.status IN ('complete', 'complete+verified')
RETURN COUNT(*);
\end{lstlisting}
	}
\end{jobQ}

\begin{jobQ}
27A
{\em 
\begin{lstlisting}[numbers=none,  showstringspaces=false,belowskip=0pt ]
MATCH  (t:title)$-$[:has_movie_info]$\rightarrow$(mi:movie_info), 
       (t:title)$-$[:movie_keyword]$\rightarrow$(k:keyword),
       (t:title)$-$[ml:movie_link]$\rightarrow$(:title), 
       (t:title)$-$[mc:movie_companies]$\rightarrow$(cn:company_name),
       (t:title)$-$[:has_complete_cast]$\rightarrow$(cc:complete_cast)
WHERE  t.production_year >= 1950 AND 
       t.production_year <= 2000 AND 
       mi.info = 'Sweden' AND k.keyword ='sequel' AND
       ml.link_type IN ('follows', 'followedBy') AND
       mc.company_type = 'production company' AND
       cn.name CONTAINS 'Film' AND
       cn.country_code <> '[pl]' AND 
       cc.subject IN ('cast', 'crew') AND 
       cc.status = 'complete'
RETURN COUNT(*);
\end{lstlisting}
}
\end{jobQ}

\begin{jobQ}
28A
{\em 
\begin{lstlisting}[numbers=none,  showstringspaces=false,belowskip=0pt ]
MATCH  (t:title)$-$[:has_movie_info]$\rightarrow$(mi:movie_info),  
       (t:title)$-$[:has_mov_info_2]$\rightarrow$(mii:mov_info_2),
       (t:title)$-$[:movie_keyword]$\rightarrow$(k:keyword), 
       (t:title)$-$[mc:movie_companies]$\rightarrow$(cn:company_name),
       (t:title)$-$[:has_complete_cast]$\rightarrow$(cc:complete_cast)
WHERE  t.production_year > 2000 AND 
       t.kind = 'movie' AND mi.info = 'Germany' AND
       mi.info_type = 'countries' AND 
       mii.info < '8.5' AND 
       mii.info_type = 'rating' AND
       k.keyword = 'murder' AND
       mc.note CONTAINS '(200' AND
       cn.country_code <> '[us]' AND
       cc.subject = 'crew' AND
       cc.status <> 'complete+verified'
RETURN COUNT(*);
\end{lstlisting}
}
\end{jobQ}

\begin{jobQ}
29A
{\em 
\begin{lstlisting}[numbers=none,  showstringspaces=false,belowskip=0pt ]
MATCH  (t:title)$-$[:has_movie_info]$\rightarrow$(mi:movie_info),  
       (t:title)$-$[:movie_keyword]$\rightarrow$(k:keyword),
       (t:title)$-$[:has_complete_cast]$\rightarrow$(cc:complete_cast), 
       (t:title)$-$[ci:cast_info]$\rightarrow$(n:name),
       (n:name)$-$[:has_aka_name]$\rightarrow$(:aka_name), 
       (n:name)$-$[:has_person_info]$\rightarrow$(pi:person_info),
       (t:title)$-$[mc:movie_companies]$\rightarrow$(cn:company_name)
WHERE  t.production_year <= 2010 AND 
       t.production_year >= 2000 AND
       t.title = 'Shrek 2' AND
       mi.info_type = 'release dates' AND 
       mi.info STARTS WITH 'Japan:' AND
       k.keyword = 'computer-animation' AND
       cc.status = 'complete+verified' AND
       cc.subject = 'crew' AND ci.role = 'actress' AND
       ci.name = 'Queen' AND 
       ci.note CONTAINS '(voice' AND n.gender = 'f' AND 
       n.name CONTAINS 'An' AND 
       pi.info_type = 'trivia' AND
       cn.country_code = '[us]'
RETURN COUNT(*);
\end{lstlisting}
}
\end{jobQ}

\begin{jobQ}
30A
{\em 
\begin{lstlisting}[numbers=none,  showstringspaces=false,belowskip=0pt ]
MATCH  (t:title)$-$[:has_movie_info]$\rightarrow$(mi:movie_info), 
       (t:title)$-$[:has_mov_info_2]$\rightarrow$(mii:mov_info_2),
       (t:title)$-$[:movie_keyword]$\rightarrow$(k:keyword), 
       (t:title)$-$[:cast_info]$\rightarrow$(n:name),
       (t:title)$-$[:has_complete_cast]$\rightarrow$(cc:complete_cast)
WHERE  t.production_year > 2000 AND 
       mi.info_type = 'genres' AND 
       mi.info = 'Horror' AND 
       mii.info_type = 'votes' AND
       k.keyword = 'murder' AND n.gender = 'm' AND 
       cc.subject IN ('cast', 'crew') AND 
       cc.status = 'complete+verified'
RETURN COUNT(*);
\end{lstlisting}
}
\end{jobQ}

\begin{jobQ}
31A
{\em 
\begin{lstlisting}[numbers=none,  showstringspaces=false,belowskip=0pt ]
MATCH  (t:title)$-$[:has_movie_info]$\rightarrow$(mi:movie_info), 
       (t:title)$-$[:has_mov_info_2]$\rightarrow$(mii:mov_info_2),
       (t:title)$-$[:movie_keyword]$\rightarrow$(k:keyword), 
       (t:title)$-$[:cast_info]$\rightarrow$(n:name),
       (t:title)$-$[:movie_companies]$\rightarrow$(:company_name)
WHERE  mi.info_type = 'genres' AND 
       mi.info = 'Horror' AND
       mii.info_type = 'votes' AND 
       k.keyword = 'murder' AND n.gender = 'm'
RETURN COUNT(*);
\end{lstlisting}
}
\end{jobQ}

\begin{jobQ}
32A
	{\em 
\begin{lstlisting}[numbers=none,  showstringspaces=false,belowskip=0pt ]
MATCH  (t:title)$-$[:movie_keyword]$\rightarrow$(k:keyword), 
       (t:title)$-$[:movie_link]$\rightarrow$(:title)
WHERE  k.keyword ='character-name-in-title'
RETURN COUNT(*);
\end{lstlisting}
	}
\end{jobQ}

\begin{jobQ}
33A
{\em 
\begin{lstlisting}[numbers=none,  showstringspaces=false,belowskip=0pt ]
MATCH  (t1:title)$-$[ml:movie_link]$\rightarrow$(t2:title), 
       (t1:title)$-$[:has_mov_info_2]$\rightarrow$(mii1:mov_info_2),
       (t2:title)$-$[:has_mov_info_2]$\rightarrow$(mii2:mov_info_2), 
       (t1:title)$-$[mc1:movie_companies]$\rightarrow$(cn1:company_name),
       (t2:title)$-$[mc2:movie_companies]$\rightarrow$(cn2:company_name)
WHERE  t1.kind = 'tv series' AND 
       ml.link_type IN ('follows', 'followedBy') AND
       t2.kind = 'tv series' AND 
       t2.production_year >= 2005 AND 
       t2.production_year <= 2008 AND 
       mii1.info_type = 'rating' AND 
       mii2.info_type = 'rating' AND 
       mii2.info < '3.0' AND 
       cn1.country_code = '[us]' AND 
RETURN COUNT(*);
\end{lstlisting}
	}
\end{jobQ}

\end{appendix}

\fi

\end{document}